\documentclass{article}
\usepackage[utf8]{inputenc}
\usepackage[letterpaper, margin=1in]{geometry}
\usepackage{cite}
\usepackage{graphicx}
\usepackage{geometry}
\usepackage[mathlines]{lineno}
\usepackage{amsmath}
\usepackage{etoolbox}  
\usepackage{amssymb}
\usepackage{xcolor}
\usepackage{subfig} 
\usepackage{float}
\usepackage[export]{adjustbox}
\usepackage{verbatim}

\newcommand*\linenomathpatch[1]{%
  \cspreto{#1}{\linenomath}%
  \cspreto{#1*}{\linenomath}%
  \csappto{end#1}{\endlinenomath}%
  \csappto{end#1*}{\endlinenomath}%
}
\newcommand*\linenomathpatchAMS[1]{%
  \cspreto{#1}{\linenomathAMS}%
  \cspreto{#1*}{\linenomathAMS}%
  \csappto{end#1}{\endlinenomath}%
  \csappto{end#1*}{\endlinenomath}%
}

\newcommand{\textcolorblue}{}

\expandafter\ifx\linenomath\linenomathWithnumbers
  \let\linenomathAMS\linenomathWithnumbers
  \patchcmd\linenomathAMS{\advance\postdisplaypenalty\linenopenalty}{}{}{}
\else
  \let\linenomathAMS\linenomathNonumbers
\fi

\linenomathpatch{equation}
\linenomathpatchAMS{gather}
\linenomathpatchAMS{multline}
\linenomathpatchAMS{align}
\linenomathpatchAMS{alignat}
\linenomathpatchAMS{flalign}

\makeatletter
\patchcmd{\mmeasure@}{\measuring@true}{
  \measuring@true
  \ifnum-\linenopenaltypar>\interdisplaylinepenalty
    \advance\interdisplaylinepenalty-\linenopenalty
  \fi
  }{}{}
\makeatother

\title{Physically-informed data-driven modeling of active nematics
}
\author{Matthew Golden, Roman O. Grigoriev, Jyothishraj Nambisan, Alberto Fernandez-Nieves}

\date{November 2021}

\begin{document}

\maketitle

\begin{abstract}
    A continuum description is essential for understanding a variety of collective phenomena in active matter. However, building quantitative continuum models of active matter from first principles can be extremely challenging due to both the gaps in our knowledge and the complicated structure of nonlinear interactions. Here we use a novel physically-informed data-driven approach to construct a complete mathematical model of an active nematic from experimental data describing kinesin-driven microtubule bundles confined to an oil-water interface. We find that the structure of the model is similar to the Leslie-Ericksen and Beris-Edwards models, but there are significant and important differences. Rather unexpectedly, elastic effects are found to play no role \textcolorblue{in the experiments considered}, with the dynamics controlled entirely by the balance between active stresses and friction stresses.
\end{abstract}

\section*{Introduction}
Active matter and the associated emergent phenomena such as spontaneous organized motion have attracted a lot of attention recently \cite{toner_tu_1998, toner2005, ramaswamy2010, marchetti2013, julicher2018}. Many different types of active matter exist at different length scales. In this paper we focus on a particular class of such systems, known as active nematics \cite{doostmohammadi2018}, which feature highly elongated apolar interacting units. Some notable examples include systems comprised of vibrated monolayers of cylindrical rods \cite{narayan2007}, microtubules \cite{sanchez2012}, actin filaments \cite{kumar2018}, and certain types of bacteria \cite{li2018} suspended in a layer of fluid. 

Active nematics exhibit a range of complex defect-mediated flows \cite{sanchez2012, wu2017, lemma2019, opathalage2019} and a number of hydrodynamic models have been proposed to understand and quantify the observed flow patterns and transitions between dynamical regimes, mostly in two spatial dimensions,  \cite{thampi2013, giomi2014, thampi2014epl, giomi2015, thampi2015, thampi2016, green2017, doostmohammadi2017, doostmohammadi2018, martinez2019, alert2020}. Most of these models are variants of the Leslie-Ericksen model \cite{ericksen1961,leslie1968} or the \textcolorblue{explicitly nematic} Beris-Edwards model \cite{degennes1993,beris1994}, which provide a coarse-grained description of {\it microscopic} nematic molecules in three spatial dimensions. Their applicability to \textit{macroscopic} filaments such as microtubules (MTs) or actin bundles is questionable, especially when those filaments are confined to an interface between a pair of immiscible fluids \cite{sanchez2012, lemma2019, opathalage2019}. Indeed, \textcolorblue{while} existing hydrodynamic models \textcolorblue{capture some aspects of the observed phenomena \cite{pearce2019, pearce2021}}, they also fail to describe a number of experimental observations for MT suspensions \cite{lemma2019, opathalage2019}.

Furthermore, hydrodynamic models of active nematics have a dozen or so parameters, few of which can be directly measured. Some progress has been made in indirectly identifying parameters in models of known functional form from experimental data for both filamentous bacteria \cite{li2018} and MT suspensions \cite{colen2021}. However, a more fundamental question of what the correct form of the hydrodynamic model of a particular active nematic system is remains unresolved. The present study goes a step further than these studies and shows that both the functional form of the model and the values of the coefficients can be identified from experimental data using an algorithm known as sparse physics-informed discovery of empiric relations (SPIDER) \cite{gurevich2021}.

Building on a body of machine learning literature devoted to spatially extended nonequilibrium systems \cite{bar1999, xu2008, rudy2017, schaeffer2017, gurevich2019, reinbold2020, reinbold2021}, SPIDER combines the relevant domain knowledge (e.g., the symmetries of the physical system) and experimental measurements (e.g., orientation and velocity fields) to identify a complete set of parsimonious physical relations necessary to describe the observed dynamics quantitatively. Unlike neural networks that can be trained to reproduce the observed dynamics, including those of active nematics \cite{colen2021, zhou2021} but yield little physical insight, SPIDER yields a set of physical relations in the familiar form of partial differential equations (PDEs) which can be directly compared against existing models. Moreover, these physical relations are easily interpretable and provide substantial new physical insight, especially when discrepancies with existing models are found. 

\section*{Problem Description}

\textcolorblue{
We consider here an active nematic where the flow is driven by a suspension of MTs confined at a water-oil interface \cite{sanchez2012}. This system has traditionally been described \cite{thampi2013,thampi2014,giomi2015,green2017,colen2021} using variations of either the Leslie-Ericksen or the Beris-Edwards model. Both models were originally derived to provide a continuum descriptions of \textcolorblue{molecular} nematic \textcolorblue{liquid crystals} in three spatial dimensions. Neither model has a first-principles justification for suspensions of \textcolorblue{colloidal mesogens} confined to a two-dimensional interface between a pair of fluid layers which are themselves in contact with one or more rigid boundaries. For instance, while both fluids are incompressible in three dimensions, the interfacial flow generally will not satisfy the divergence-free conditions in two spatial dimensions as is assumed by both models. Similarly, the viscous flows in the two fluid layers can be very different from those in a three-dimensional volume. Some attempts have been made to address vertical confinement by, e.g., introducing Rayleigh friction \cite{thampi2014,doostmohammadi2016}. However, this does not address, but rather exacerbates, the key problem: both models contain a multitude of material parameters which cannot all be computed or measured, making direct quantitative comparison between experiment and theory a \textcolorblue{challenging task}.
}

\textcolorblue{
First-principles analyses uniformly assume, but never prove, that the system can be described by a two-dimensional model, whatever its functional form. While confinement indeed effectively constrains the motion of MTs to two spatial dimensions, the flow in both fluid layers driven by that motion generally remains fully three-dimensional \cite{martinez2021}. One of the key objectives of this study is therefore to \textcolorblue{address whether} three-dimensional effects are weak and the system affords an effective quantitative description in two spatial dimensions, analogous to the treatment of other weakly turbulent flows in thin fluid layers \cite{tithof2017}. We will assume that such a model can be synthesized using measurements of three physical observables at the two-dimensional oil-water interface. These observables --  the director field, the flow velocity, and a local order parameter -- appear in most first-principles models of active nematics.
}

\textcolorblue{
In particular, the Leslie-Ericksen model is formulated in terms of the director field ${\bf n}$. Nematic symmetry ${\bf n} \rightarrow -{\bf n}$ combined the presence of topological defects, which is a generic feature of this system, implies that the director field cannot be defined globally as a continuous field.  
To avoid this complication, most theoretical models instead use the globally continuous nematic tensor $Q_{ij} = S n_i n_j$, where $0 \leq S \leq 1$ is a scalar order parameter, or its traceless counterpart $\bar{Q}_{ij}$. The scalar order parameter $S$ measures the degree of local alignment of nematic molecules and mainly serves to describe disorder arising from thermal motion of microscopic nematic molecules. In contrast, \textcolorblue{away from defects}, MTs tend to be \textcolorblue{well}-aligned due to a combination of their large aspect ratio and their relatively strong interaction; \textcolorblue{hence, we set $S=1$ in our analysis.}
}

\textcolorblue{
Unlike \textcolorblue{molecular nematics}, which tend to have a uniform density, MT\textcolorblue{-based systems} are nonuniform, with their density or, more accurately, packing fraction $\phi$ varying between zero in the neighborhood of topological defects and unity far from the defects. The density field $\phi$ plays a role analogous -- but note equivalent -- to the (dis)order parameter $S$. Note that the values of $\phi\approx 1$ can only be achieved due the near-perfect alignment of the MTs. While some phenomenological models that include the density field have been proposed \cite{giomi2014,mitchell2021}, they lack proper justification or validation. Most commonly, $\phi$ is simply assumed to be a constant in the first-principles models of MT suspensions.
}

\section*{Results}
\subsection*{Data-driven model discovery}

The difficulties facing first-principles analyses call for an alternative approach where a quantitative model of a particular experimental system is constructed directly from experimental data. Here we will rely on a recently introduced technique named sparse physically-inspired discovery of empiric relations (SPIDER) which has already been validated using both numerical data describing a highly turbulent flow driven by pressure gradients \cite{gurevich2021} and experimental data describing a weakly turbulent flow driven by Lorentz force \cite{reinbold2021}. The latter case in particular has numerous similarities with the problem considered here. Most notably, it also involves two thin fluid layers supported by a rigid bottom boundary and strong vertical confinement is used to synthesize a quantitative two-dimensional model of the nominally three-dimensional fluid flow.

\begin{figure}[h!]
    \begin{tabular}{@{}cc@{}}
    \centering
    \subfloat[]{%
        \includegraphics[width=0.25\textwidth]{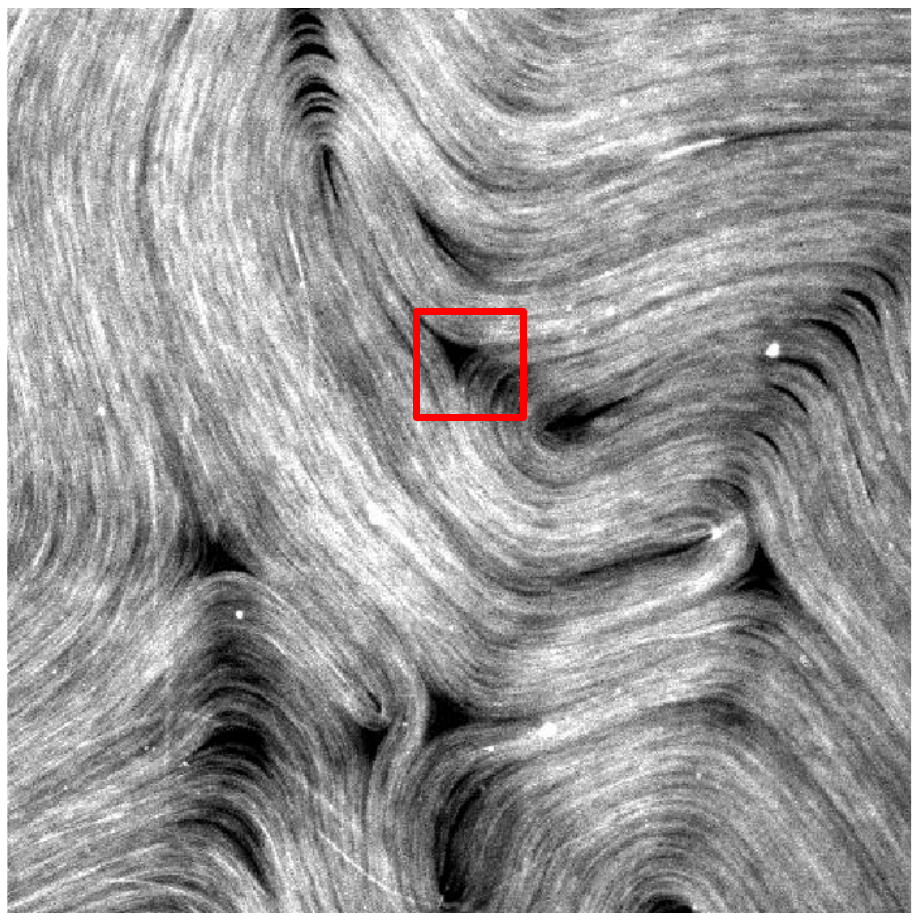}%
        \label{fig:first}
    }&
    \subfloat[]{%
        \includegraphics[width=0.25\textwidth]{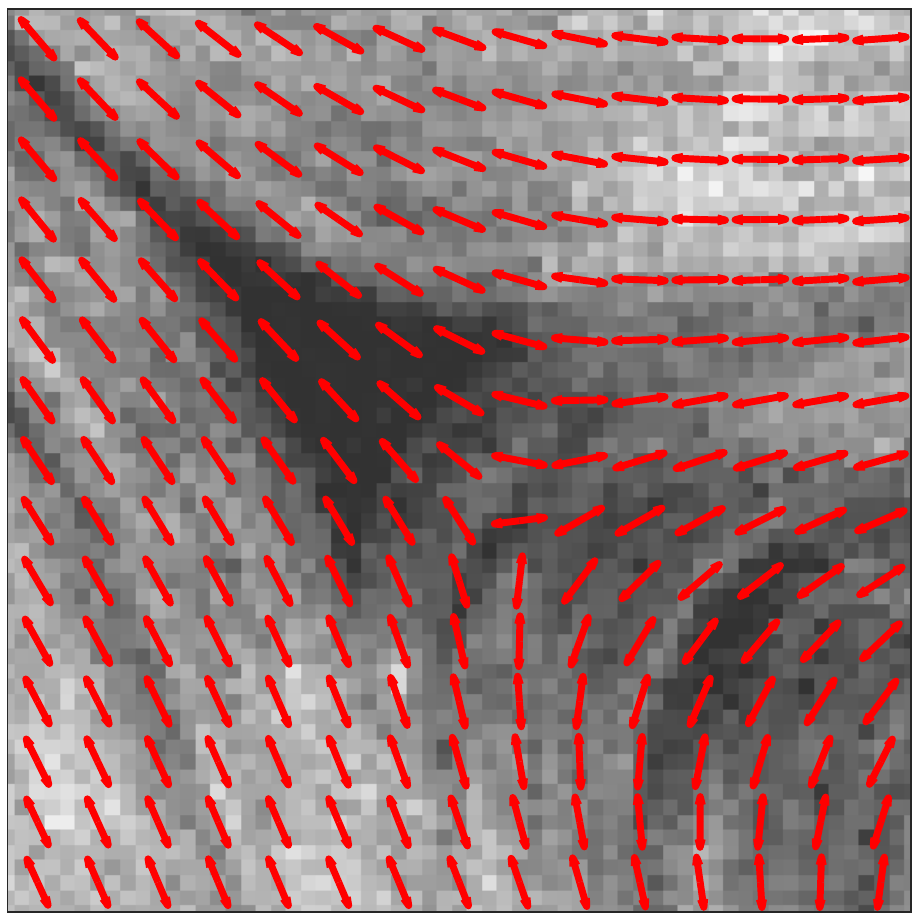}%
        \label{fig:zoomed_defect}
    }\\
    \subfloat[]{%
        \includegraphics[width=0.25\textwidth]{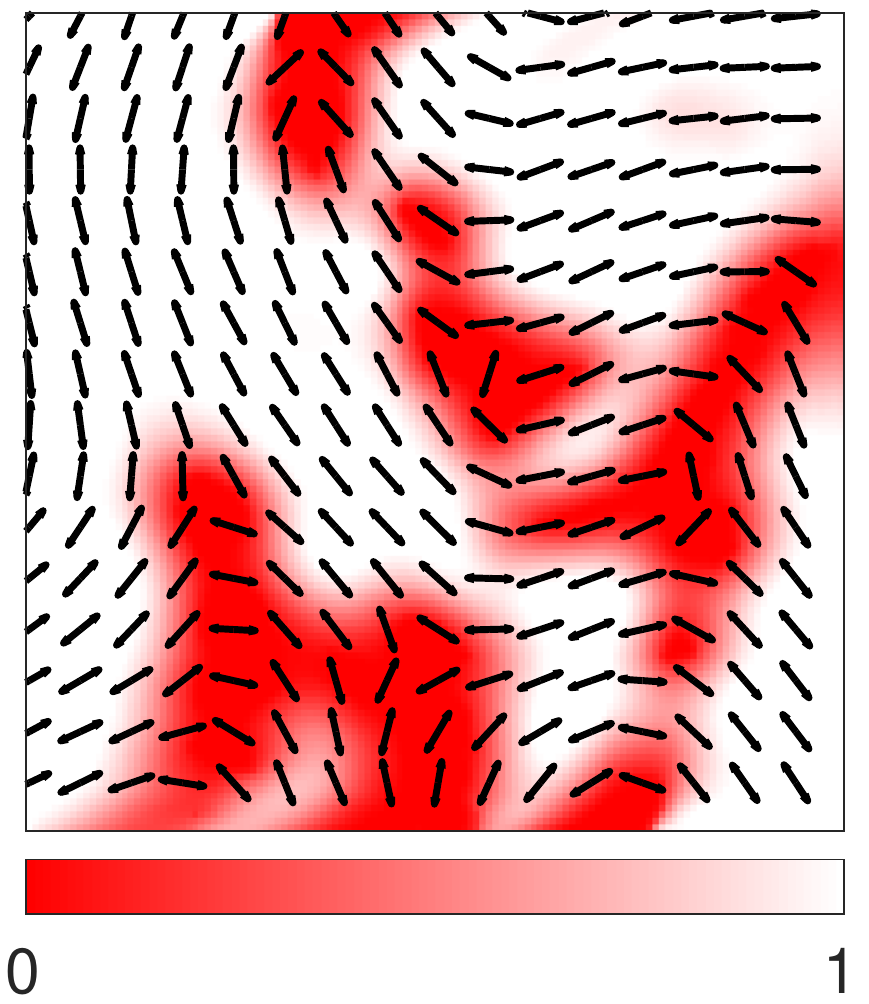}%
        \label{fig:zoomed_defect}
    }&
    \subfloat[]{%
        \includegraphics[width=0.25\textwidth]{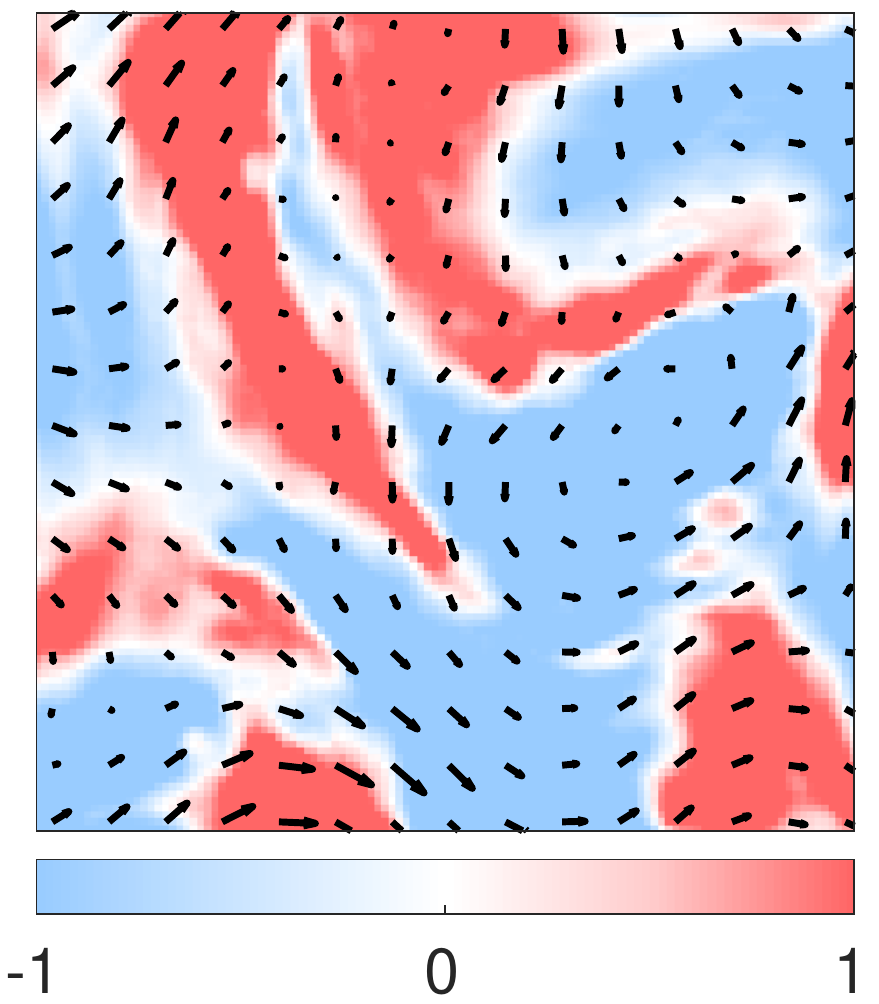}%
        \label{fig:flow_velocity}
    }\\
    
    \end{tabular}\hfill
    \begin{minipage}{0.4\textwidth}
    \caption{ \textbf{Raw experimental images and the extracted fields.} (a) A snapshot of the MTs. The complete image is shown, with the red box highlighting a small region centered on a $-1/2$ topological defect. Panel (b) shows the zoomed-in view of the red box. The extracted director field ${\bf n}$ (red arrows) does not line up with the orientation of the MTs in the center of the image, which indicates that director field data is unreliable near topological defects. (c) Director field (black arrows) and the mask $\psi$ (color), (d) The flow field ${\bf u}$ (black arrows) and the corresponding vorticity $\omega = -2\Omega_{xy}$ (color). Panels (c) and (d) show the vector fields corresponding to panel (a) on a much coarser grid than that on which the data are available.}      
    \label{fig:data}
    \end{minipage}
\end{figure}

Before discussing the application of SPIDER to the experimental system considered here, let us make several observations regarding the data, which are represented by sequences of snapshots of dense flourescently labeled MT bundles, such as the one shown in Figure \ref{fig:data}(a). From these one can identify the director field ${\bf n}$, the flow velocity ${\bf u}$, and the MT density $\phi$, all at the interface on a rectangular grid spanned by the two spatial coordinates and time. Examples of reconstructed fields are shown in Figure \ref{fig:data}(b-d). The spatial resolution of the images however is insufficient to resolve the fast variation of $\phi$ and ${\bf n}$ near the topological defects. Moreover, no reliable information about either ${\bf n}$ or ${\bf u}$ can be obtained in the regions where $\phi\approx 0$ (dark areas in the image) which typically surround topological defects, as illustrated in Figure \ref{fig:data}(b). As a result, we exclude these regions from our analysis. Outside of those regions, $\phi\approx 1$ is effectively constant. Therefore, only the fields ${\bf n}$ and ${\bf u}$ serve as useful data. 

\begin{figure}[t!]
    \centering 
    \includegraphics[width = 1\textwidth]{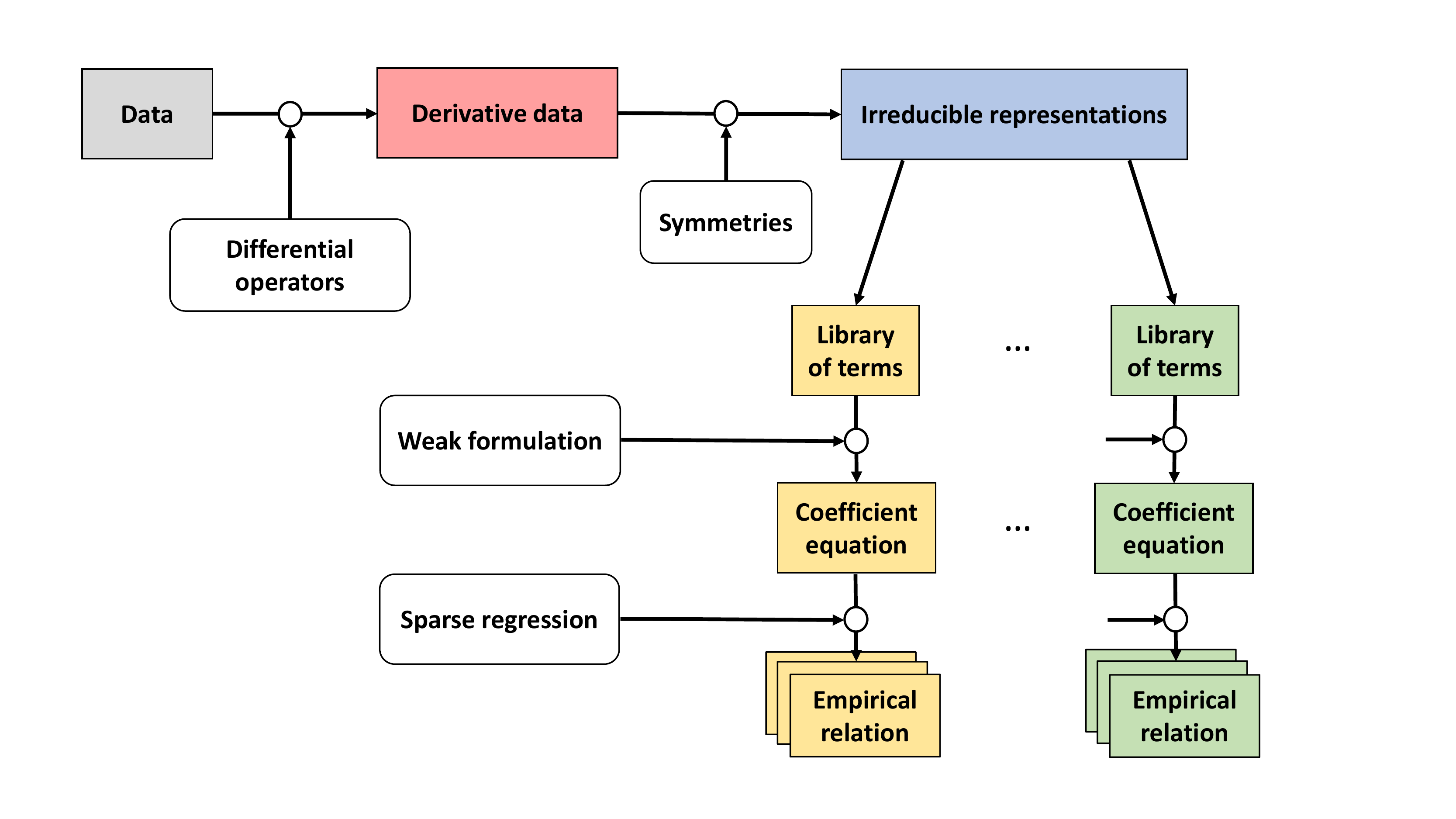}
    \caption{\textcolorblue{
    \textbf{A schematic representation of the SPIDER algorithm.} Tensors $F^r$ are constructed in symbolic form from fields and their derivatives and projected into irreducible representations of the underlying symmetry group, yielding a set of libraries. Weak form of the corresponding equation \eqref{eq:lib} evaluated using appropriately sampled data yields a coefficient equation \eqref{eq:coeff}. Finally, a sparse regression algorithm is applied to each coefficient equation to identify one or more empirical relations.
    }
    }
    \label{fig:SPIDER}
\end{figure}

These velocity and director fields are, in fact, sufficient to \textcolorblue{synthesize a hydrodynamic model of this experimental system} with the help of SPIDER, as described in detail in the Methods section. \textcolorblue{Figure \ref{fig:SPIDER} summarizes the key ingredients and steps of the algorithm. First, a sufficiently large set of tensor products $F^r({\bf n},{\bf u})$, up to rank 2, is constructed in symbolic form from the vector fields ${\bf n}$ and ${\bf u}$ as well as their spatial and temporal derivatives. These are split into irreducible components according to the symmetries of the system, yielding libraries of terms with similar transformation laws.} For instance, terms that are invariant with respect to the nematic symmetry and transform as vectors under rotation form one library, while terms that are invariant under rotations but change sign when ${\bf n}$ is replaced with $-{\bf n}$ form another. Each library represents a PDE of the form
\begin{align}\label{eq:lib}
    \sum_rc_rF^r=0
\end{align}
with coefficients $c_r$ that are assumed to be constant, \textcolorblue{reflecting the symmetry of the problem with respect to translations in space and time}. Note that the vertical confinement implies that any model that might be discovered, would be two-dimensional, so 
$\nabla=\hat{\bf x}\nabla_x+\hat{\bf y}\nabla_y$, ${\bf u}=\hat{\bf x}u_x+\hat{\bf y}u_y$, and so on.

\textcolorblue{To combat measurement noise \cite{gurevich2019}, PDE \eqref{eq:lib} is converted to its weak form. Specifically, it is multiplied by one of several smooth weight functions and integrated} over one of many rectangular spatiotemporal domains to generate an overdetermined linear system of equations
\begin{align}\label{eq:coeff}
    G{\bf c}=0
\end{align}
for the coefficients $c_r$. \textcolorblue{The matrix $G$ contains integrals of individual terms which are evaluated using properly nondimensionalized data. This process is repeated for each library, yielding a set of coefficient equations \eqref{eq:coeff}.} To make sure that regions with unreliable director and velocity field data are excluded, the weights are constructed as a product with a mask $\psi$ that vanishes in the regions to be excluded. In practice, the mask is constructed automatically based on the values of $\phi$ and $\nabla_in_j$. An example of a mask overlaid on the director field is shown in Figure \ref{fig:data}(c) and corresponding Supplementary Movie S4, which shows how the mask evolves in time.

\begin{figure}[h!]
    \centering 
    \includegraphics[width = 0.75\textwidth]{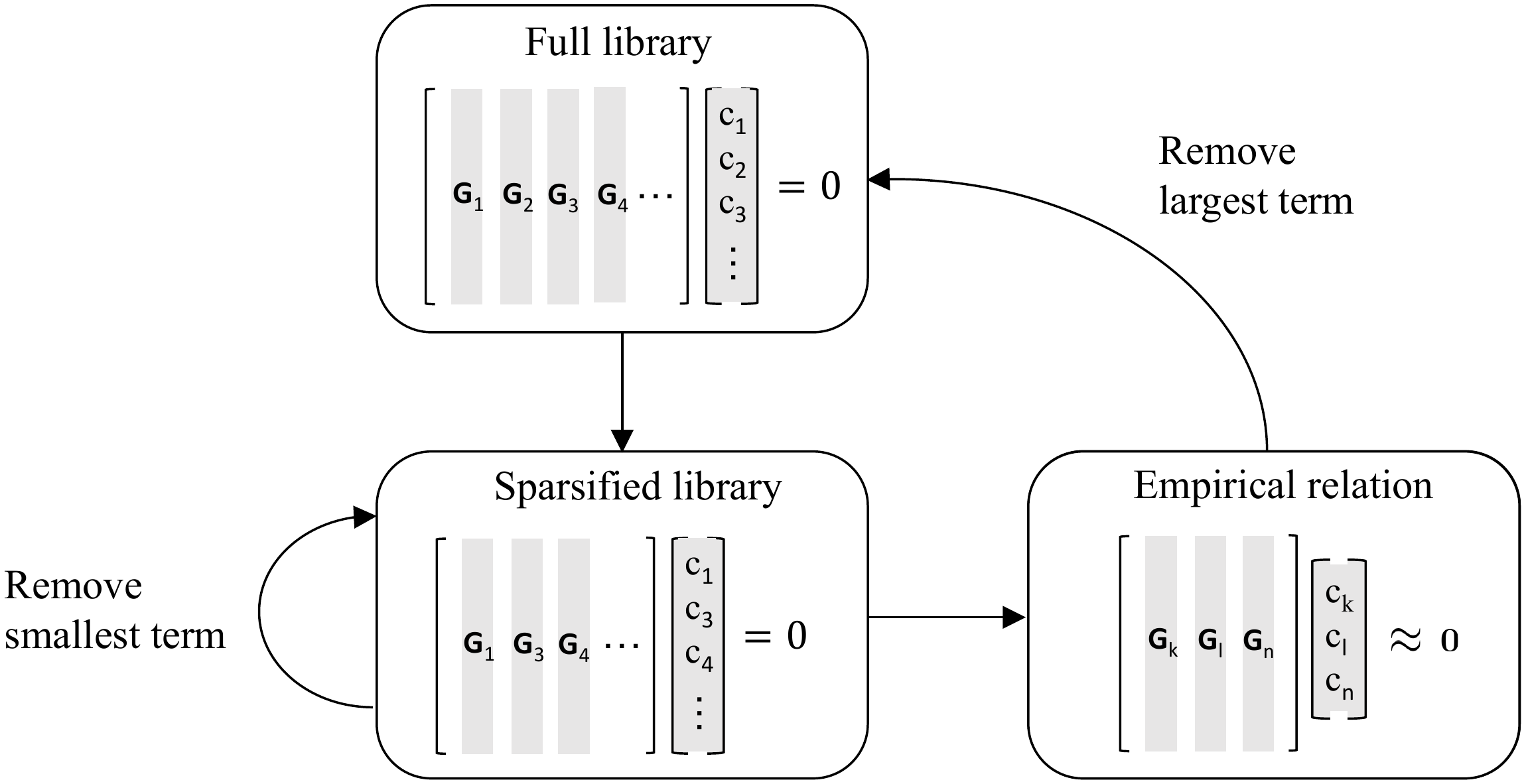}
    \caption{\textcolorblue{ \textbf{An iterative regression algorithm for finding a set of sparse solutions for the coefficient equation \eqref{eq:coeff}.} An empirical relation balancing sparsity and accuracy is identified using sequential regression of the full library. The largest term in that relation is then removed from the full library and the procedure is repeated to find additional empirical relations contained in the library. ${\bf G}_n$ represents the $n$-th column of the matrix $G$.}}
    \label{fig:sparse_regression}
\end{figure}

\textcolorblue{Finally, a sparse regression algorithm illustrated by Figure \ref{fig:sparse_regression} is applied to each coefficient equation \eqref{eq:coeff}. This iterative algorithm finds one or more approximate sparse solutions ${\bf c}$, which correspond to parsimonious yet quantitatively accurate relations between the fields ${\bf u}$ and ${\bf n}$. In this manner, nine equations were found across six different libraries (see Supplementary Material).} It is straightforward to show that all of them can be derived from the following set of three fundamental relations: an incompressibility condition
\begin{align}
   \nabla_i u_i = 0, \label{eq:incomp}
\end{align}
an evolution equation for the director field
\begin{align}
   \partial_t n_i + c_1 u_j \nabla_j n_i + c_2 \Omega_{ij} n_j + c_3\hat{P}_\perp \bar{A}_{ij} n_j = 0\label{eq:ndot},
\end{align}
and a stress balance 
\begin{align}
    \bar{A}_{kl} \bar{Q}_{kl} \bar{Q}_{ij} + c_5 \bar{Q}_{ij} = 0, \label{eq:stress_balance}
\end{align}
with the coefficients $c_1=(0.99\pm0.8\%)$, $c_2=(-0.95 \pm 0.7\%)$, $c_3=(-0.95 \pm 1\%)$, and $c_5=(-0.56 \pm 1\%)$. 
\textcolorblue{Here $\hat{P}_\perp$ is the projection operator onto the direction normal to ${\bf n}$, $A_{ij}$ and $\Omega_{ij}$ represent, respectively, the symmetric and antisymmetric components of \textcolorblue{the velocity gradient} tensor $\nabla_iu_j$, and the bar denotes the trace-free component of a symmetric tensor. The two terms in equation \eqref{eq:stress_balance} could be interpreted as an anisotropic viscous stress $\sigma^v_{ij}=\mu\bar{Q}_{ij} \bar{A}_{kl} \bar{Q}_{kl}$ and an active stress $\sigma^a_{ij}=\alpha\bar{Q}_{ij}$, with the coefficient $c_5\propto-\alpha/\mu$ relating the strengths of activity and viscosity. The activity coefficient $\alpha$ is positive (negative) for an extensile (contractile) nematic \cite{aditi_2002, green2017}. The coefficient $c_5$ is indeed negative, suggesting that $\alpha>0$, as it should be for an extensile nematic considered here.}

\begin{figure}[H]
    \begin{tabular}{@{}cc@{}}
    \centering
    \subfloat[]{%
        \includegraphics[width=0.24\textwidth,valign=t]{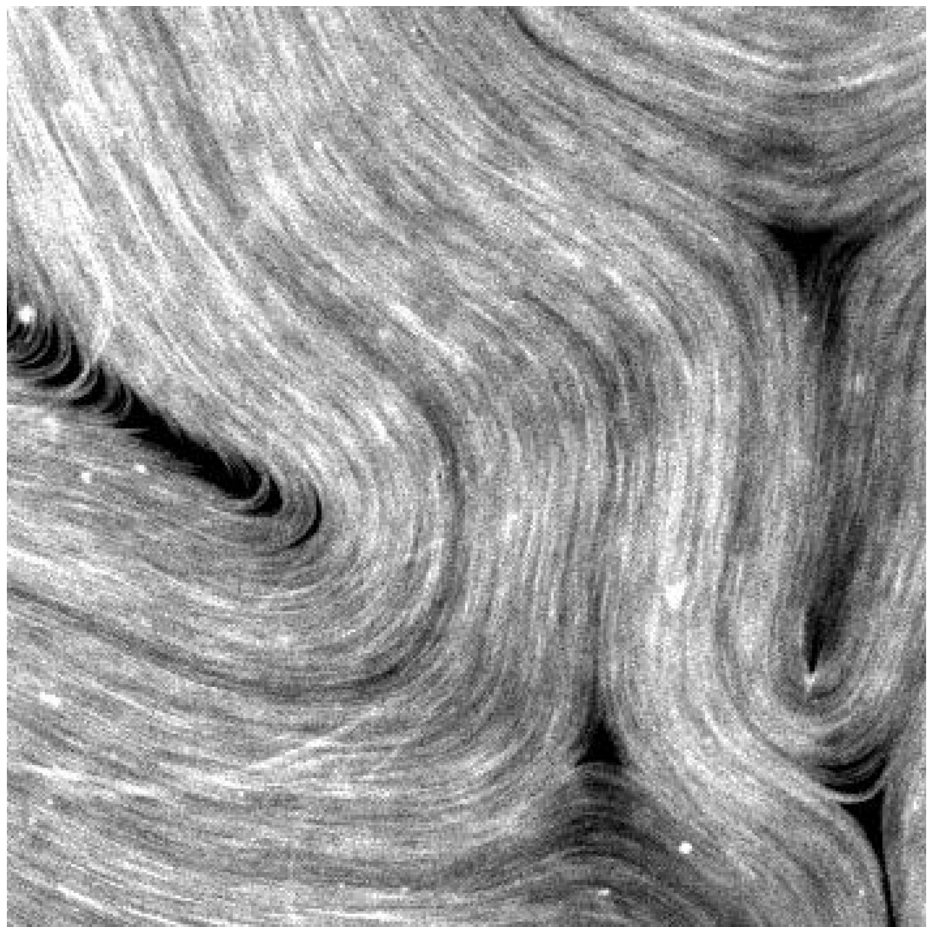}%
        \vphantom{\includegraphics[width=0.24\textwidth,valign=t]{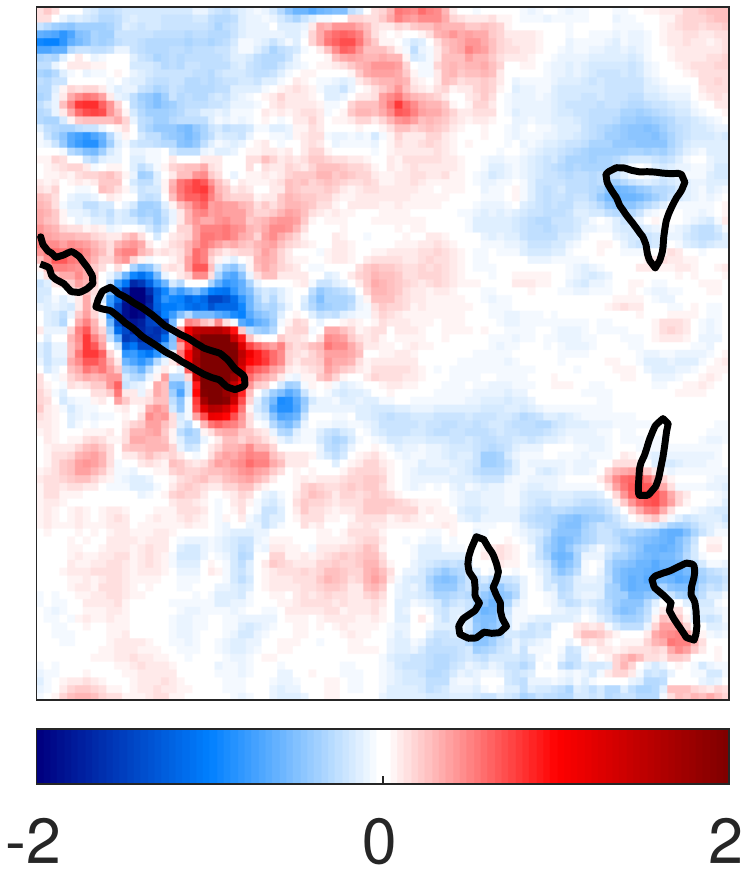}}%
        \label{fig:cropped_experiment_for_residuals}
    }&
    \subfloat[]{%
        \includegraphics[width=0.25\textwidth, valign=t]{figures/div_u.pdf}%
        \label{fig:zeta1}
    }\vspace{2mm}
    \\
     \subfloat[]{%
        \includegraphics[width=0.25\textwidth]{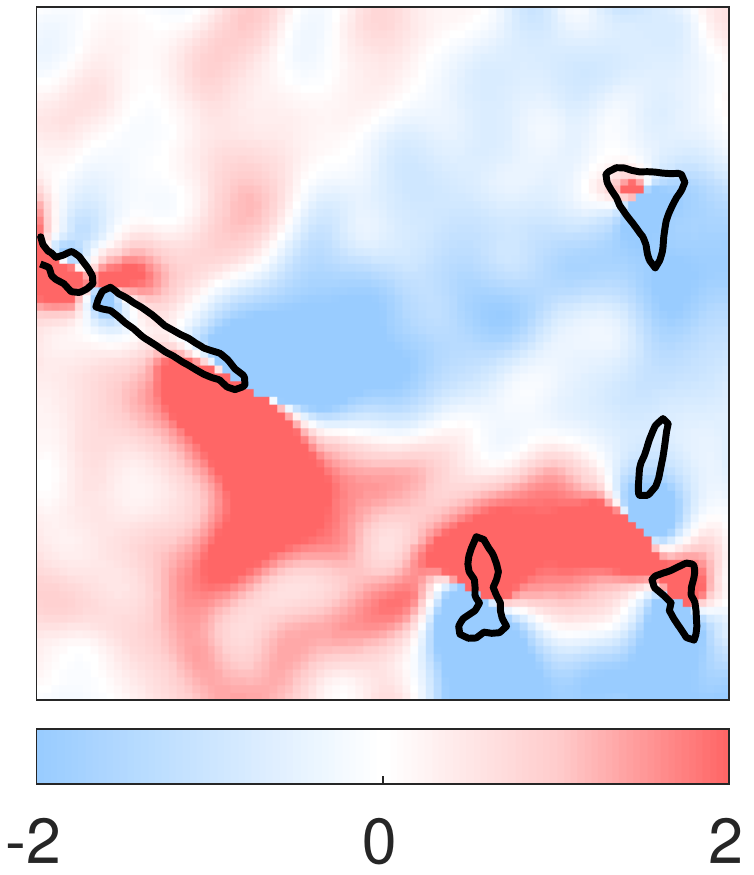}%
        \label{fig:nematodynamics_1}
    }&
    \subfloat[]{%
        \includegraphics[width=0.25\textwidth]{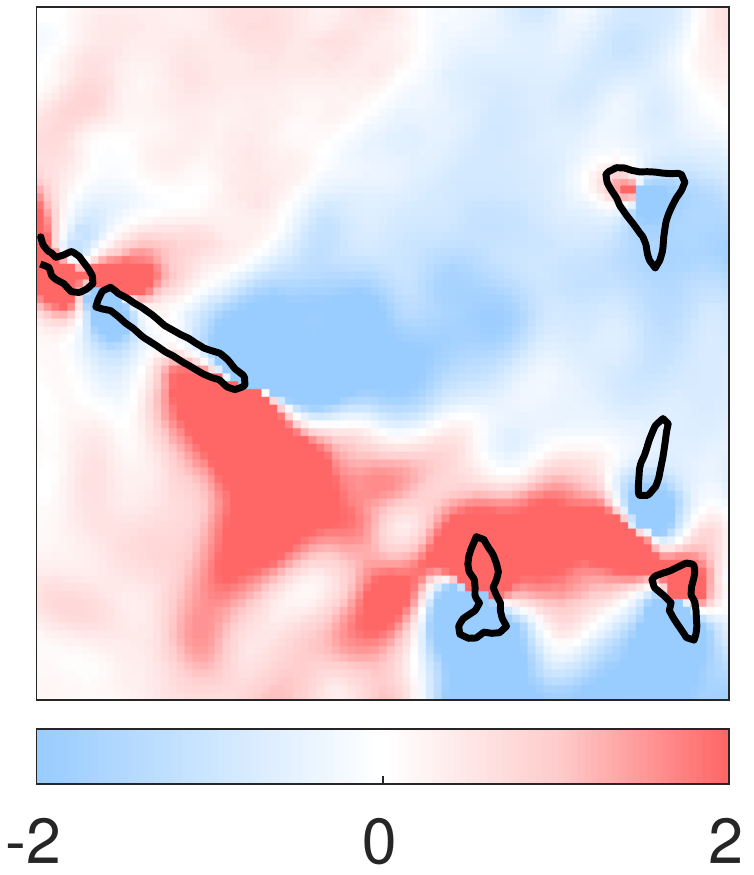}%
        \label{fig:nematodynamics_2}
    }\vspace{2mm}
    \\
    \subfloat[]{%
        \includegraphics[width=0.25\textwidth]{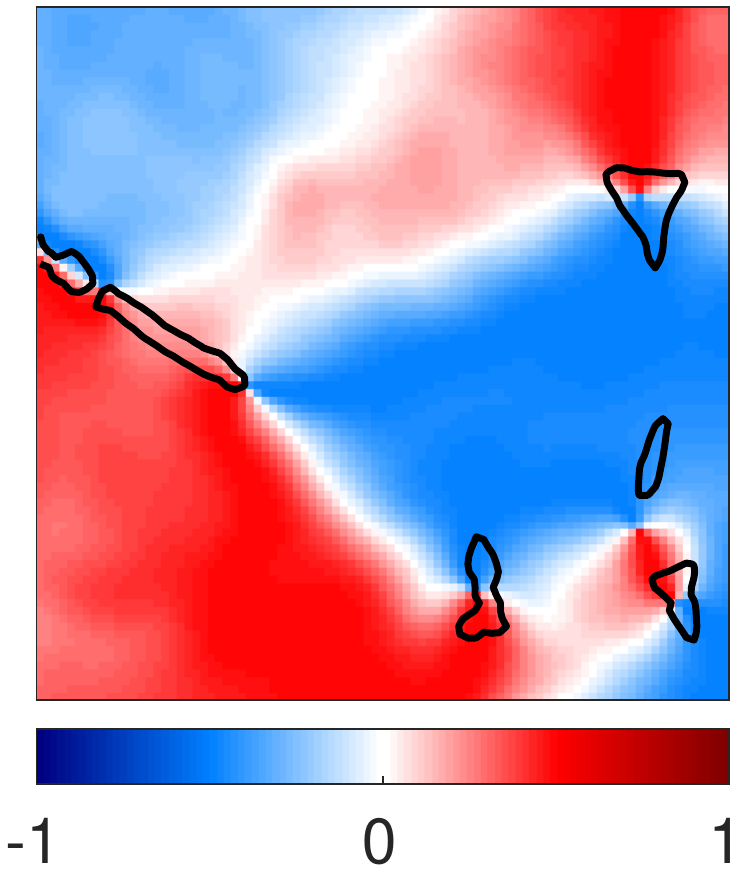}%
        \label{fig:Q_11}
    }&
    \subfloat[]{%
        \includegraphics[width=0.25\textwidth]{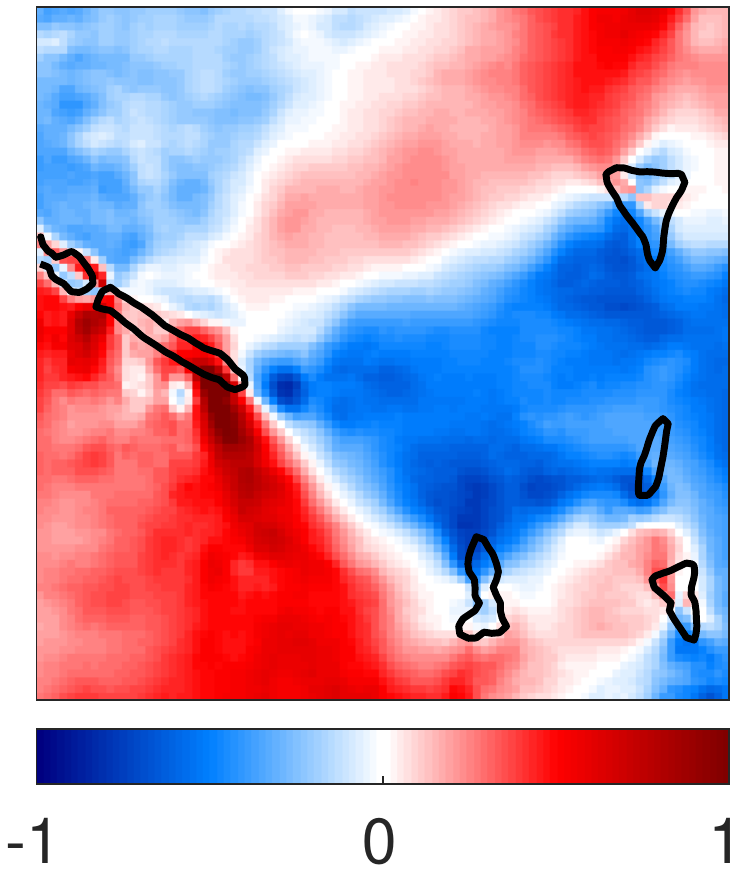}%
        \label{fig:diss_11}
    }\vspace{2mm}
    \\
    \subfloat[]{%
        \includegraphics[width=0.25\textwidth]{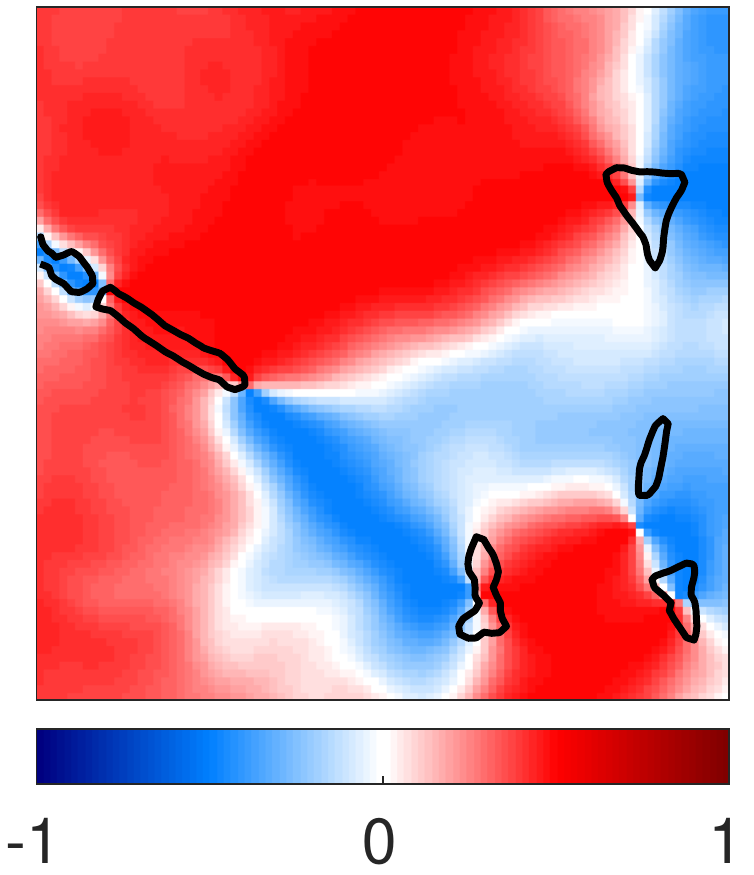}%
        \label{fig:Q_12}
    }&
    \subfloat[]{%
        \includegraphics[width=0.25\textwidth]{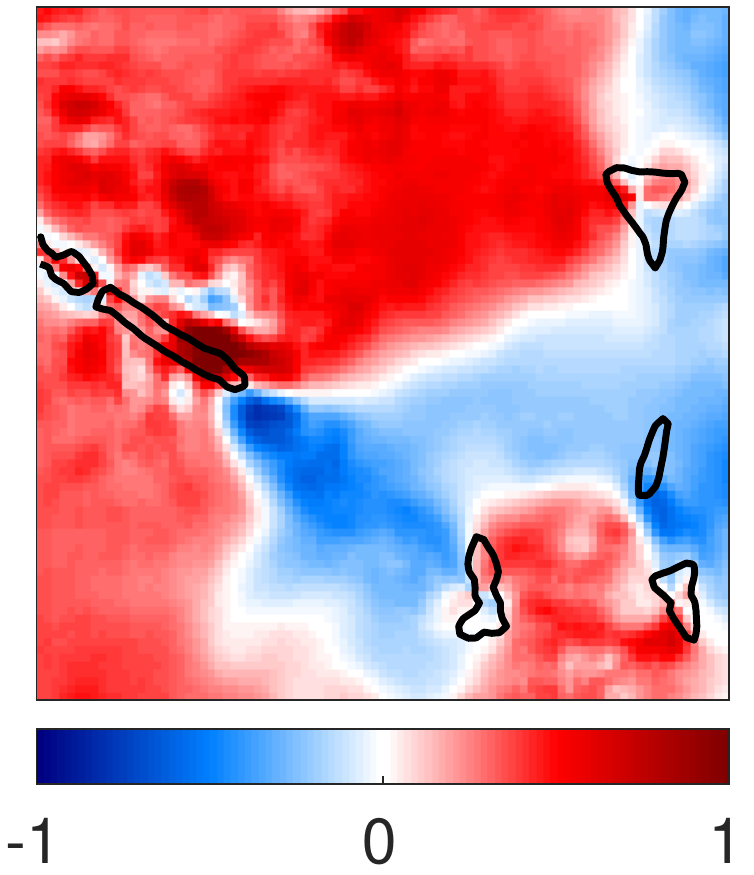}%
        \label{fig:diss_12}
    }

    \end{tabular}\hfill
    \begin{minipage}{0.4\textwidth}
    \caption{\textbf{The strong form of the identified relations.} Symbolic regression identifies physical relations in weak form. To check the validity of the corresponding PDEs in strong form, we computed each term on the entire spatial domain using finite differences. Panel (a) shows a cropped snapshot of the MTs, and all other panels correspond to this snapshot. Panel (b) shows the divergence $\nabla_i u_i$ of the interfacial flow. Panel (c) shows the observed angular velocity $\partial_t \theta=\varepsilon_{ij} n_i \partial_t n_j$ of the MTs and panel (d) its value \textcolorblue{reconstructed} using the vector relation \eqref{eq:ndot}. The remaining panels compare the two components of the active and viscous stresses in arbitrary units: the diagonal component $\sigma^a_{11}$ (e) and $-\sigma^v_{11}$ (f) and the off-diagonal component $\sigma^a_{12}$ (g) and $-\sigma^v_{12}$ (h). The viscous stress shown in panels (f) and (g) involves spatial derivatives and is therefore much noisier than the active stress shown in panels (e) and (g). Solid black curves in panels (b-h) correspond to a level set of the number density field $\phi$ and describe the edges of the regions devoid of MTs.}      
    \label{fig:residuals}
    \end{minipage}
\end{figure}

Before discussing how these PDEs are related to the existing models of active nematics, let us emphasize that SPIDER ensures that these relations are satisfied in weak form. To check that the strong form of these PDEs is also satisfied, we evaluated and compared various terms at every location in space and time. In particular, the divergence $\nabla\cdot{\bf u}$ for a typical snapshot is shown in Figure \ref{fig:residuals}(b) and Supplementary Movie S1. We find that its magnitude is small almost everywhere, \textcolorblue{consistent with what has been previously reported for the active nematic under consideration \cite{pearce2021}.} Regions where the divergence takes large, positive or negative, values are collocated with the regions where the MT density is low ($\phi\approx 0$). These regions are excluded from our analysis, so the velocity field can be considered essentially divergence-free where the MTs are dense ($\phi\approx 1$).

Figures \ref{fig:residuals}(c,d) and Supplementary Movie S2 compare the \textcolorblue{field $\partial_t{\bf n}$ computed directly from the data with that given by equation \eqref{eq:ndot}.} In this case, we find good agreement in the entire domain, not just in the regions with $\phi\approx 1$. Finally, Figure \ref{fig:residuals}(e-h) and Supplementary Movie S3 compare the active stress $\sigma^a_{ij}$ and the viscous stress $\sigma^v_{ij}$ that appear in the stress balance relation \eqref{eq:stress_balance}. The two independent components of the (symmetric) stress tensor are shown in Figure \ref{fig:residuals}(e-h) and Supplementary Movie S3. Again, we find good agreement in the entire domain. The minor discrepancies that can be seen are a result of insufficient accuracy in the numerical evaluation of the derivatives of the velocity field.

\subsection*{Discussion}

Let us now turn to the discussion of the physical insight that the identified relations suggest. First of all, it should be emphasized that we obtained a {\it complete} mathematical description of the problem in the regions with nearly uniform density $\phi$ of MTs. This description has the same number of equations (three) as both the Leslie-Ericksen and the Beris-Edwards models, describing the fluid flow and the orientation of the MTs.
Two of these relations, the incompressibility condition \eqref{eq:incomp} and the evolution equation for the director field \eqref{eq:ndot}, are the same as in the Leslie-Ericksen model
\textcolorblue{
\begin{subequations}\label{eq:LE}
\begin{align}
    \label{eq:LEa} 
    \partial_t{\bf n}+{\bf u}\cdot\nabla {\bf n}&
    =\Omega{\bf n}+\lambda \hat{P}_\perp A {\bf n} +\Gamma\frac{\delta}{\delta {\bf n}} F[{\bf n}],\\
    \label{eq:LEc} 
    \nabla\cdot{\bf u}&=0,\\
    \label{eq:LEb} 
    \partial_t{\bf u}+{\bf u}\cdot\nabla {\bf u}&=\rho^{-1}\nabla\cdot\sigma,
\end{align}
\end{subequations}
}
since the coefficients $c_1$  and $c_2$ are both very close to the expected values of $\pm1$. Furthermore, $\lambda=-c_3$ is found to be very close to unity, as expected for thin filaments \cite{larson1999}. \textcolorblue{It is notable that no contributions from the free energy $F[{\bf n}]$, including those due to elasticity, are identified in the evolution equation \eqref{eq:ndot}.}

\textcolorblue{The remaining equation \eqref{eq:stress_balance} however is rather unexpected.} This is a tensor relation representing local balance between active and viscous stresses, not a vector relation representing momentum balance, as is the case in the Leslie-Ericksen and the Beris-Edwards model. \textcolorblue{Given that we are dealing with a creeping flow, it is hardly surprising that the terms $\partial_t{\bf u}$ and ${\bf u}\cdot\nabla{\bf u}$ representing inertia can be ignored, so that equation \eqref{eq:LEb} should reduce} to the force balance
\begin{align}
    \nabla\cdot\sigma=0.
\end{align}
This equation is consistent with the discovered relation \eqref{eq:stress_balance}, provided the stress tensor can be decomposed as $\sigma=\sigma^v+\sigma^a$, where the viscous stress and active stress were defined previously.  Note that, again, no elastic contribution is found, which is consistent with the absence of elastic effects in the evolution equation \eqref{eq:ndot}. 
\textcolorblue{In our experimental setup, which is characterized by a relatively low density of topological defects,} it is the balance of active and viscous stresses that controls the flow \cite{thampi2015,thampi2016,doostmohammadi2018} rather than the balance between active and elastic stresses, as is more commonly assumed \cite{thampi2013, giomi2014, thampi2014epl, giomi2015, green2017, doostmohammadi2017, lemma2019, martinez2019, alert2020}. In addition, we find that the pressure is essentially constant and thus, we can neglect it as well. Finally, the stress tensor $\sigma^v$ representing viscous effects is highly anisotropic, in contrast to what is commonly assumed. \textcolorblue{It represents a special case of the more general phenomenological expression
\begin{align}\label{eq:stress_Leslie}
    \sigma_{ij}^v=\nu_4A_{ij}+\beta_1(A_{kl}n_kn_l)n_in_j
    +\beta_2(A_{ik}n_kn_j+A_{jk}n_kn_i)
\end{align}
proposed by Leslie \cite{leslie1968}.
Here $\beta_1=\nu_1-\lambda(\nu_2+\nu_3)$, $\beta_2=\nu_5+\lambda\nu_2$, and $\nu_1$ through $\nu_5$ are the ``Leslie viscosities.'' Only one term is} present ($\beta_1\ne 0$), the other two are either absent or too small to be detected ($\nu_4=\beta_2=0$).


While we refer to the two-dimensional tensors $\sigma^a$ and $\sigma^v$ as ``stress tensors'' in keeping with the convention, it is important to understand that their components do not represent the stress according to its standard definition in three dimensions. For instance, the forces at the interface generated by kinesin (active stress) are balanced by the $xz$ and $yz$ components of the viscous stresses in the fluid layers above and below the interface which do not appear in our effectively two-dimensional description. Rather, it is more appropriate to think of the relation \eqref{eq:stress_balance} as a two-dimensional ``projection'' of the proper stress balance relation in three dimensions. In particular, although $\sigma^v$ involves spatial derivatives of the interfacial flow velocity, rather than the flow velocity itself, it does represent friction stresses \cite{thampi2014,doostmohammadi2016}, as explained in the Supplementary Material. One can also find there a discussion of the correct physical interpretation of the parameter $c_5$.


Although the evolution equation for the \textcolorblue{$Q$-tensor} contained in the Beris-Edwards model is not listed among the three fundamental relations \eqref{eq:incomp}-\eqref{eq:stress_balance}, it follows immediately from the evolution equation \eqref{eq:ndot} if the latter is multiplied by $n_j$. It is also identified via symbolic regression with coefficients very close to those in equation \eqref{eq:ndot} (see Supplementary Material). In this case, as before, no elastic (or, more generally, free energy) contributions are found. Symbolic regression also identifies several other relations that follow from one of the fundamental relations. For instance, we find a simple relation  
\begin{align}\label{eq:stress_balance_scalar}
    \bar{Q}_{ij}\bar{A}_{ij}+c'_5=0
\end{align}  
with $c'_5 = (-0.55 \pm 0.3\%)\approx c_5$ that is equivalent to the stress balance \eqref{eq:stress_balance}, although it does not allow an equally intuitive physical interpretation. 


Note that two of the three physical relations \eqref{eq:incomp}-\eqref{eq:stress_balance} describing this system involve no time derivatives and cannot be identified using methods such as SINDy \cite{brunton2016} which assume their presence. Physical constraints play a crucial role in constructing the libraries, and one should be careful to include as much physics as possible and, at the same time, avoid using assumptions that only appear logical but are not physically grounded. In addition, when properly constrained, symbolic regression becomes an extremely powerful and general tool for synthesizing new scientific knowledge, as the results presented here vividly illustrate. 


\subsection*{Limitations and future work}

It is worth reiterating that the model we have identified does not provide a full description of the dense MT suspension at a flat interface. This model only describes regions where the curvature of the \textcolorblue{MTs} is low and their density $\phi$ is high and nearly uniform. The dynamics in this system are \textcolorblue{controlled} by the topological defects, the neighborhoods of which have been excluded in our analysis. \textcolorblue{To properly account for these dynamics,} our model has to be generalized to describe the regions around the defects where the curvature is high and the density $\phi$ varies in both space and time. That generalization has to include an evolution equation for $\phi$ and incorporate the dependence on $\phi$ into the remaining governing equations. \textcolorblue{The former is undoubtedly the continuity equation reflecting mass conservation
\begin{align}
    \partial_t\phi+\nabla\cdot{\bf j}=0.
\end{align}
Diffusion of MTs is negligible due to both confinement to the interface and their large size, hence the flux is likely dominated by advection \cite{mitchell2021}, ${\bf j}=\phi{\bf u}$, although curvature corrections \cite{giomi2014} are possible.}

\textcolorblue{
The dimensional version of the model \eqref{eq:incomp}-\eqref{eq:stress_balance} contains only one parameter $c_5$ which defines a time scale. 
\textcolorblue{It does not contain any parameters that can be used to define a length scale.}
The absence of elastic stresses in our model suggests that interaction between topological defects is mediated by the flow in the two fluid layers when the mean defect separation $L$ is large compared with the layer thickness $h$. Indeed, in our experiment, $L\approx 240$ $\mu$m is large\textcolorblue{r than} $h=50$ $\mu$m. Our results, however, do not exclude the possibility that elastic effects might play a role in the high-curvature regions, requiring generalization of the model.} Equation \eqref{eq:ndot} does not appear to need any modification, as it describes the evolution of the director field quite accurately in regions even with low $\phi$, as illustrated by Figure \ref{fig:residuals}(c,d) and the corresponding Supplementary Movie S2. On the other hand, both relations describing the fluid flow have to be generalized. 

In particular, there is no physical reason for the interfacial fluid flow to be divergence-free everywhere in two dimensions, even though the data suggests that the incompressibility equation \eqref{eq:incomp} is satisfied away from the defects. As illustrated by Figure \ref{fig:residuals}(a,b) and the corresponding Supplementary Movie S2, $\nabla\cdot{\bf u}$ is large and positive in the neighborhood of topological defects with +1/2 charge where $\phi$ changes between low and high values. The regions with $\nabla\cdot{\bf u}$ large and negative are also collocated with the regions where $\phi$ varies between low and high values, but represents {\it past} locations of these moving topological defects. In contrast, the flow is found to be essentially divergence-free in the neighborhood of 
topological defects with -1/2 charge.

\textcolorblue{Positive divergence is only found in high-curvature regions, suggesting that elastic effects play a key role in creating the defects and pushing the MTs apart, lowering the density $\phi$. Indeed, there is experimental evidence supporting this role \cite{opathalage2019}. We find the divergence to be negative in regions where the curvature is low and the density $\phi$ is below unity suggesting that depletion interaction plays an important role. Therefore, one might expect to see two additional dimensional parameters characterizing, respectively, the stiffness of the MT \textcolorblue{bundles} and the depletion interaction in the generalized model. In particular, the stiffness parameter can be used to define a characteristic length scale, as discussed in the Supplementary Material. Indeed, exponential distributions of vortex areas at high densities of topological defects suggest the presence of a length scale which also depends on activity \cite{lemma2019} and viscosity \cite{martinez2021}.}

The stress balance equation \eqref{eq:stress_balance} should also be generalized. In regions without MTs, where $\phi=0$, the active stresses are expected to vanish, while the viscous stresses are expected to become isotropic, with the latter balanced by the pressure gradient. Regions where the pressure gradients are nonnegligible likely correspond to locations where there is a strong flow towards or away from the interface associated with the large (positive or negative) values of $\nabla\cdot{\bf u}$.  Indeed, the largest discrepancy between the \textcolorblue{active and viscous stresses is localized to those regions as well, as illustrated by Figure \ref{fig:residuals}(e-h) and movie S2.}

\begin{figure}[H]
    \centering
    \subfloat[]{%
        \includegraphics[width=0.25\textwidth, valign=t]{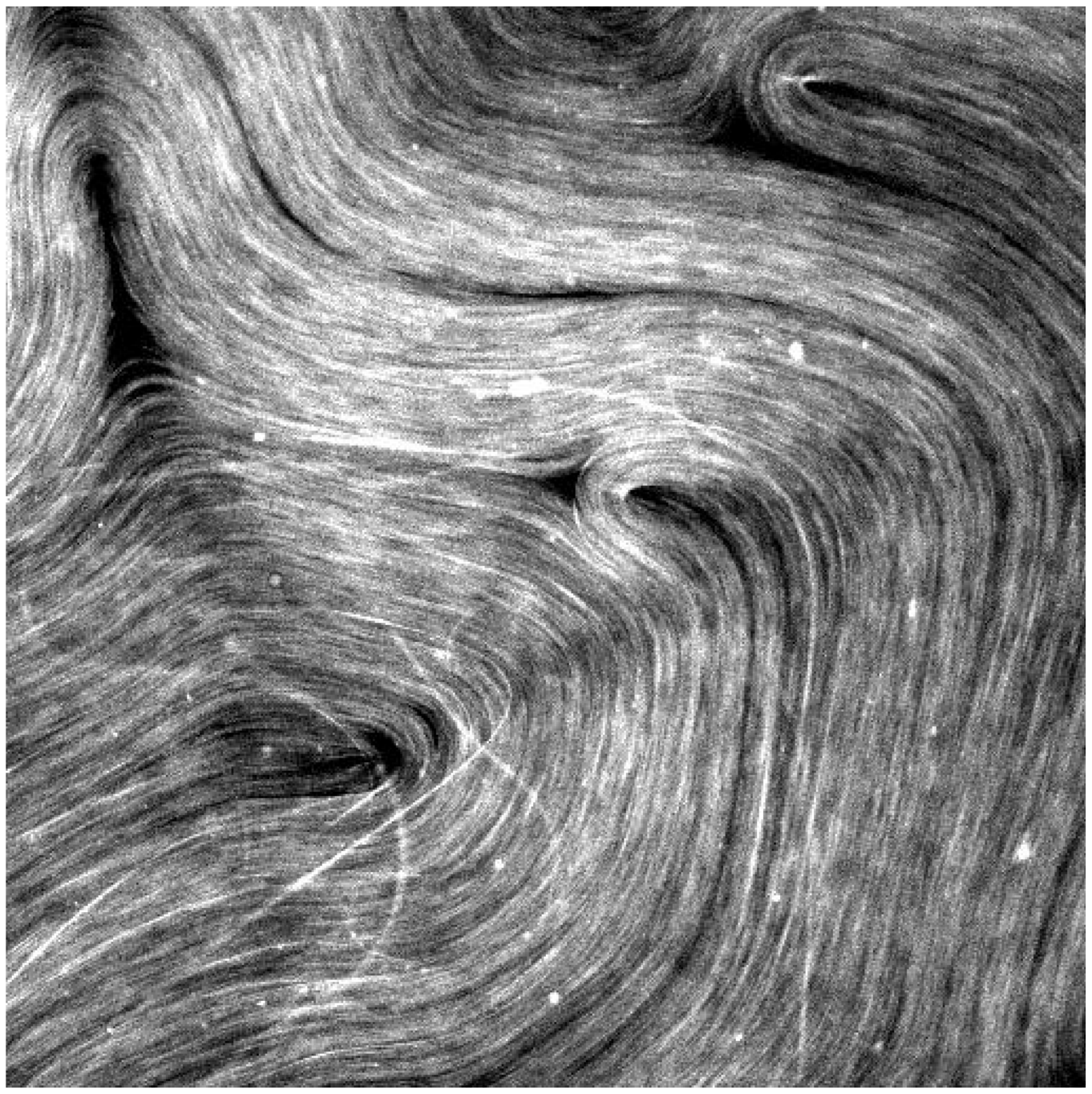}%
       \vphantom{\includegraphics[width=0.25\textwidth,valign=t]{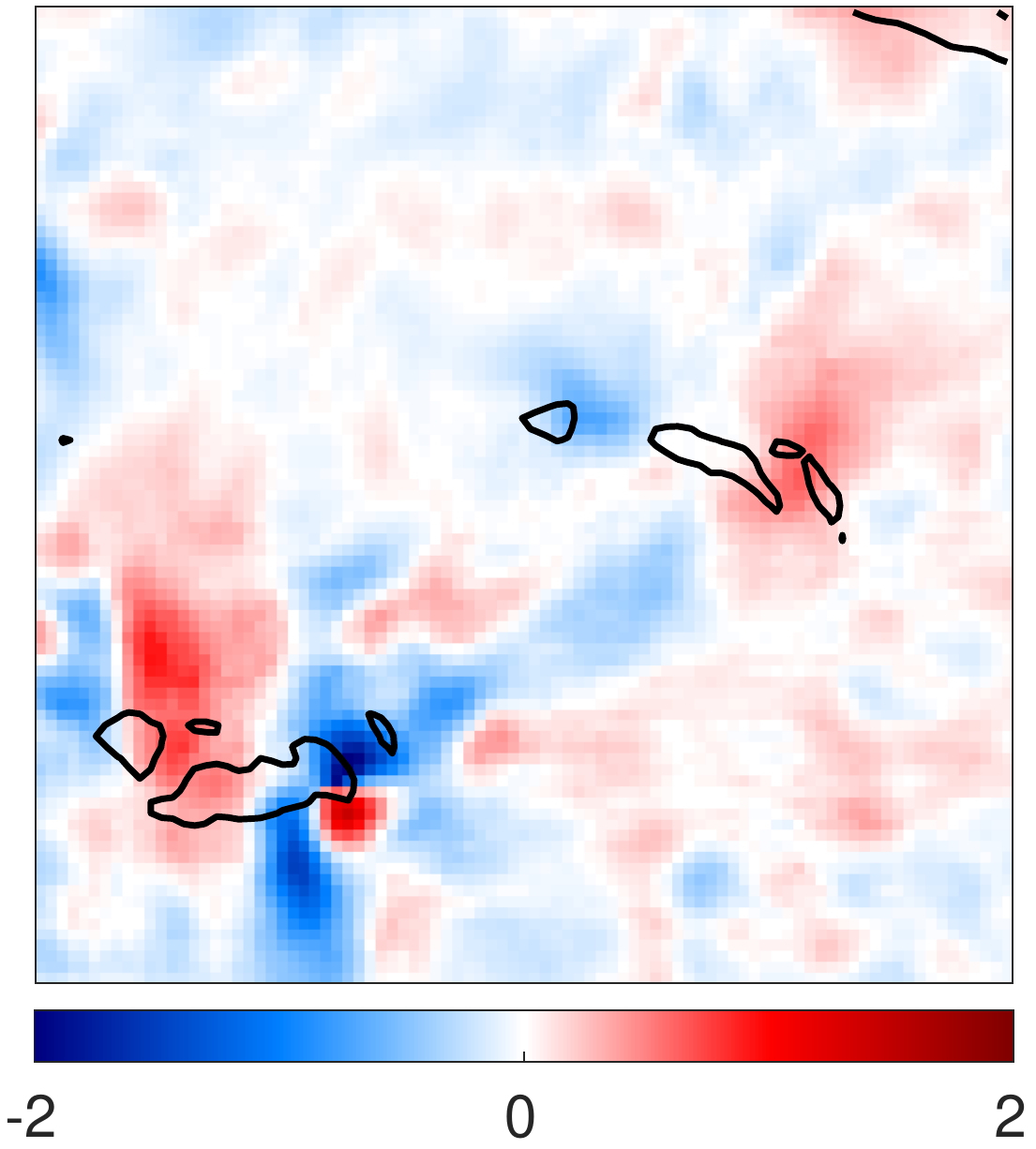}}%
        \label{fig:OOP_experiment}
    }\,
    \subfloat[]{%
        \includegraphics[width=0.25\textwidth, valign=t]{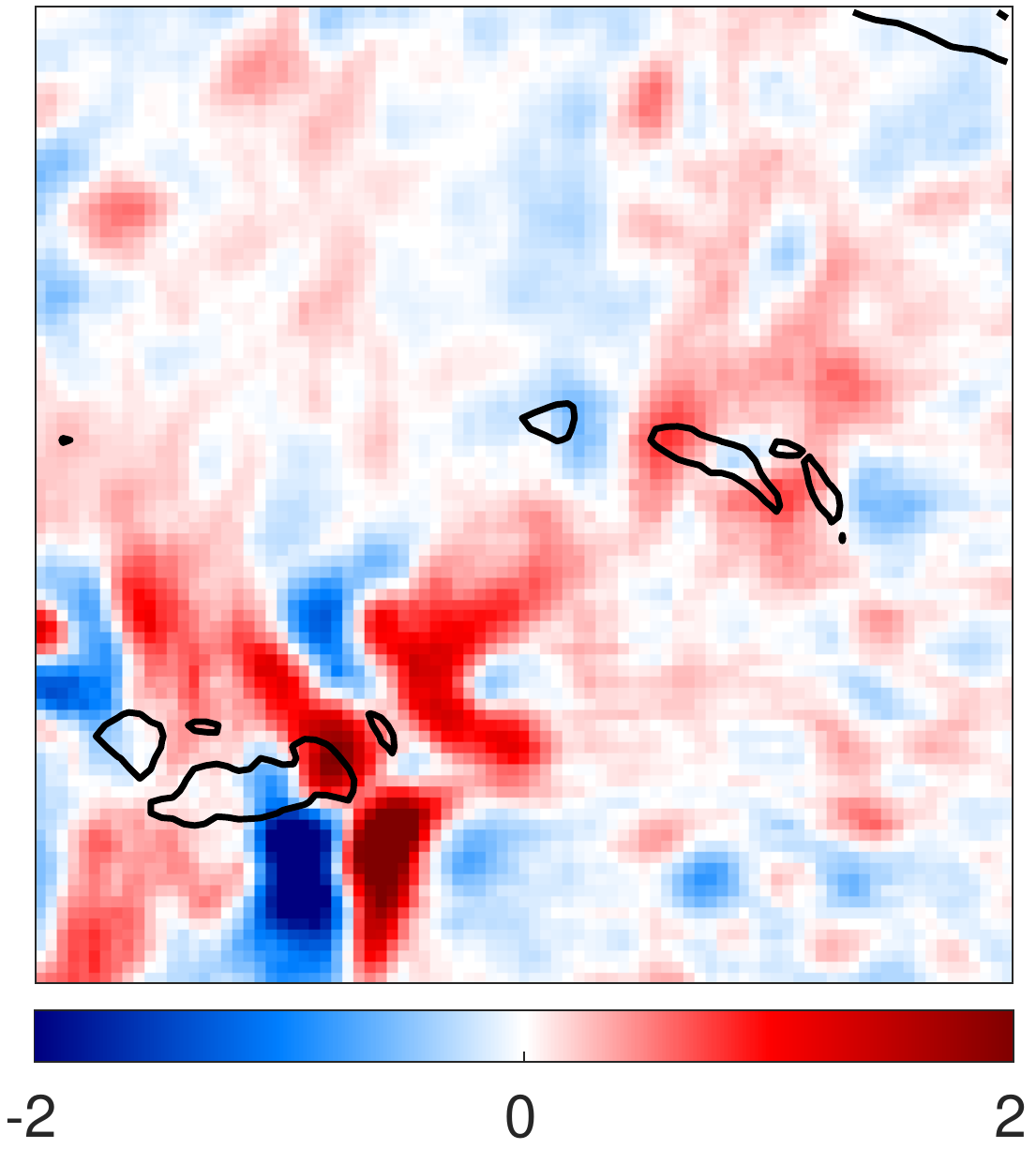}%
        \label{fig:OOP_incomp}
    }\,
    \subfloat[]{%
        \includegraphics[width=0.25\textwidth, valign=t]{figures/OOP_stress_balance_scalar.pdf}%
        \label{fig:OOP_stress_balance}
    }
    \caption{ \textbf{MT alignment and the accuracy of the identified relations.} (a) Three noticeable out-of-plane MT bundles are misaligned with the rest of the MTs at the bottom left of the image. (b) Divergence of the flow is the largest in the regions of misalignment. (c) The error (residual) of the scalar form \eqref{eq:stress_balance_scalar} of the stress balance is also the largest there.}
    \label{fig:OOP}
\end{figure}

Finally, let us revisit the assumption of the local order in the average MT orientation that underlies the validity of the hydrodynamic description of this system. While the vast majority of MTs are oriented in the same direction ($S=1$), there are rare exceptions. An example is shown in Figure \ref{fig:OOP}(a) which features several MT bundles that are misaligned with the rest. In regions where MTs cross, their orientation cannot be described by a continuous field and the hydrodynamic description breaks down. This is illustrated in Figure \ref{fig:OOP}(b,c) and Supplementary Movies S1 and S3, which show that the error in both the incompressibility condition \eqref{eq:incomp} and the stress balance \eqref{eq:stress_balance} is the largest in the region where misaligned MTs are found. This ultimately reflects that there is a 3D aspect of these MT suspensions that is often neglected.

It is straightforward to extend our analysis to include additional physical fields such as the MT density and kinesin or ATP concentration. The main challenge is the capability to measure and/or reconstruct the corresponding quantities in experiment in parallel with the director and flow fields. Additional fields need to be similarly well-resolved in space and time, so that the corresponding derivatives and integrals can be computed with reasonable accuracy.


\section*{Methods
}
\subsection*{Experimental setup and data acquisition}

\textcolorblue{We conduct our experiments on the microtubule-kinesin active nematic system pioneered in \cite{sanchez2012}.} The long rod-like MTs are bundled together via depletion interactions and are driven out of equilibrium by the action of kinesin-streptavidin motor protein complexes, which are units that induce relative motion utilizing ATP as the energy source. Depletion forces also aid in driving the MT bundles to form bundles to the oil-water fluid interface, where they execute self-sustained bending and buckling instabilities. \textcolorblue{The system is extensile, which means that active stresses cause the MT bundles to extend in length and contract in width.}  

To investigate the dynamics of defects in 2D flat space, we prepare the active nematic in a flow-cell setup where the entire pool of ingredients is confined in a 2D sealed cell roughly $10$ cm$^{2}$ in area and \textcolorblue{100 $\mu$m in thickness.} The lower surface of the cell is subjected to hydrophobic treatment (using Aquapel) and the upper surface to hydrophilic treatment (using polyacrylamide coating) to enhance wetting by the respective fluid phases. A fluorinated oil (HFE-7500 with surfactant E2K0660) forms the oil-phase, and the active MT suspension forms the water-phase. \textcolorblue{We obtained purified tubulin monomers and kinesin–streptavidin motor protein complexes from the Dogic Group at Brandeis University\cite{sanchez2012,decamp2015}. The polymerization of tubulin to MTs is performed in our lab before mixing with other biomaterials as per the protocols described in previous works \cite{sanchez2012,decamp2015,ellis2018,pearce2021}.} The final active mix has $20\%$ MTs by volume aided with $144 \mu M$ ATP. The entire flow cell is sealed by epoxy resin and centrifuged at 1000 RPM to accelerate the depletion mechanism to the interface.

We use confocal fluorescence microscopy for visualization. The MTs are labeled with AlexaFlour 647 dye and illuminated at 633 nm; the excitation and emission peaks are at 651 nm and 667 nm, respectively. After sample preparation and centrifugation, we wait for 15-20 minutes to allow for uniform depletion, and then image at a constant framerate till the activity ceases. Typically, the MTs stay active for 6+ hours. Imaging is done using $10\times$ and $20\times$ objectives to focus on regions with area on the order of mm$^{2}$, away from the edges of the flow cell. The imaging process results in a time series of 8-bit grayscale images, which are stored as the raw data \cite{pearce2021}.

\subsection*{Data processing}

\textcolorblue{Two fields are extracted directly from experimental data: the flow velocity ${\bf u}$ and the nematic director field {\bf n}. The video analyzed has $O(1000)$ frames with 512x512 resolution and side length 484.35 $\mu$m, collected at a frame rate of $0.88$ s${}^{-1}$.} The flow field ${\bf u}$ is extracted with Particle Image Velocimetry (PIV) performed by LaVision DaVis 10.1. The resulting flow field is only resolved on a 128x128 mesh, and all other fields are restricted to this resolution. 

\textcolorblue{The director field ${\bf n}$ is extracted using Coherence Enhanced Diffusion Filtering (CEDF)\cite{weickert1999}. CEDF determines the direction along which the spatial intensity variation is the smallest, which corresponds to local average alignment of the MTs. This allows constructing the nematic tensor order parameter $\bar{Q}_{ij}$, which upon diagonalization yields the nematic director ($\textbf{n}$).
A detailed description of the image analysis technique can be found in \cite{ellis2020}.} While this method reliably finds ${\bf n}$ relatively far from defects where MTs are dense, \textcolorblue{the low MT density} near the defects \textcolorblue{prevents resolving the director field in these regions, even when carefully using the blurring schemes inherent in our CEDF methodologies.} This can be seen in Figure \ref{fig:zoomed_defect}. 

The MT density $\phi$ can also be extracted from experimental images as a function of blurred intensity. Far from defects, the MTs can be reasonably assumed to be of uniform density. In the neighborhood of defects, the density is discontinuous; it vanishes close to defects where the director field becomes undefined. We restrict our study to the former regions, characterized by negligible density fluctuations. We also ignore the scalar order parameter $S$ which describes the local alignment of the MTs. Far from defects, $S=1$ as the MTs are all well-aligned with rare exceptions.

Both ${\bf u}$ and ${\bf n}$ fields are smoothed with a moving least squares multivariate fit, and any necessary derivatives are computed with second order centered differences. A good choice of units is helpful for model discovery, and we choose a characteristic length and time scale so that mean velocity and vorticity are both unity, $\langle |{\bf u}| \rangle = \langle |\omega| \rangle= 1$. In these units, the experimental image sequence has dimensions of $L_x\times L_y\times L_t=5\times 5\times 17$.


\subsection*{Model libraries}

Locality and smoothness imply that a physical relation between the two fields, ${\bf u}$ and ${\bf n}$, can be written as a superposition \eqref{eq:lib} of a number of terms $F^r$, each constructed from ${\bf u}$, ${\bf n}$ and/or their spatial and temporal derivatives, i.e., that relation has the form of a PDE. For instance, every relation in the Leslie-Ericksen model \eqref{eq:LE} has just such a form. For relations involving more than one term, all terms should transform in the same way under every operation in the symmetry group describing the problem \cite{gurevich2021}, which includes rotations around the vertical axis and nematic symmetry ${\bf n} \rightarrow - {\bf n}$. Hence, terms with different transformation properties are grouped into separate libraries $\{F^r\}$. 

The rotation symmetry implies that every term in a library has to transform as a tensor of a specific rank (we only consider tensors of rank 0 through 2). The nematic symmetry implies that every term in a library must involve either even or odd powers of ${\bf n}$. An arbitrary rank-2 tensor can be decomposed into a symmetric and antisymmetric part. The symmetric part can be further decomposed into a traceless component and the trace, with the latter transforming as a scalar. Therefore, without loss of generality, the rank-2 tensor library can be split into two parts: symmetric traceless (denoted with a bar) and antisymmetric (denoted with a tilde). 

All the libraries considered in this study are summarized in Table \ref{tab:library_notation} with the terms contained in various libraries listed in the Supplementary Material. Subscripts denote the tensor indices and superscripts denote the hyperparameters associated with regression, e.g., the index of the term in the library.
\begin{table}[]
    \centering
    \begin{tabular}{|c|c|c|}
        \hline
        Libraries & even powers of {\bf n} & odd powers of {\bf n} \\
        \hline
        rank-0 tensor (scalar) & $F^r$   & $\hat{F}^r$  \\
        rank-1 tensor (vector) & $F^r_i$ & $\hat{F}^r_i$\\
        symmetric traceless rank-2 tensor & $\bar{F}^r_{ij}$ & not studied \\
        antisymmetric rank-2 tensor & $\tilde{F}^r_{ij}$ & not studied \\
        \hline
    \end{tabular}
    \caption{ \textbf{The summary of the model libraries and their symmetry properties.}}
    \label{tab:library_notation}
\end{table}


\subsection*{Weak formulation}

Once a library has been constructed, the corresponding PDE \eqref{eq:lib} is converted to an over-determined system of linear algebraic equations $G{\bf c}=0$ for the unknown coefficients ${\bf c}=(c_1,c_2,\cdots)$ following Ref. \cite{reinbold2020}. This is accomplished by multiplying every term by weight functions $w_k$ and integrating the result over a rectangular spatiotemporal subdomain $V_l$, with each combination of $k$ and $l$ defining one or more rows of the feature matrix $G$. \textcolorblue{Weak form of the PDEs allows SPIDER to deal with noise levels as high as 100\% \cite{gurevich2021}.}

The weight functions $w_k$ are constructed as a product of three components: (i) an envelope $(1-x^2)^\tau(1-y^2)^\tau(1-t^2)^\tau$ which vanishes, along with $\tau-1$ derivatives, on the boundary of subdomain $V_l$ to eliminate the boundary terms left after integration by parts; (ii) a modulation term which is taken to be one of $\{1, \cos(\pi x-\theta^k_x), \cos(\pi y-\theta^k_y), \cos(\pi t-\theta^k_t)\}$ with arbitrary phases $\theta^k_m$, and (i) a mask $\psi$ which excludes unreliable data. Note that, in the above expressions,  $x$, $y$, and $t$ have been shifted and scaled such that $V_l$ becomes a cube $[-1,1]\times[-1,1]\times[-1,1]$. The choice of $\tau$ is determined by the highest order derivative that appears in the PDE \eqref{eq:lib}; here we take $\tau=4$ \cite{gurevich2019}.
The smooth mask $\psi$ vanishes in regions where the density of MTs is low and/or their curvature is high and approaches unity far from those regions. It is constructed by successively smoothing the Boolean field $\psi_0$, which vanishes when the image intensity is below some threshold or derivatives of ${\bf n}$ are above some threshold and is unity otherwise.

After performing the integration by parts to move as many derivatives as possible from the library term containing noisy data onto a smooth weight function (see Supplementary Material for details), each integral is evaluated numerically using the trapezoidal rule \cite{gurevich2019}. The subdomains $V_l$ are chosen to include sufficiently many grid points in every direction to ensure reasonable accuracy of numerical quadrature (54x54x65 grid points) and are centered at uniformly distributed random points of the spatiotemporal domain describing the data set in order to ensure the data is well sampled and to avoid linear dependence.

\subsection*{Sparse regression}

Once the feature matrix $G$ has been constructed, we look for a sparse solution ${\bf c}$ that corresponds to a parsimonious physical relation that balances simplicity and accuracy. Simplicity can be measured in a number of ways; here we take it to be determined by the number of nonzero coefficients $c_r$. The accuracy can be quantified by a properly normalized residual; we choose to use the 2-norm of the residual for the weak form of the relation, $\eta =\Xi^{-1} \| G {\bf c} \|_2$. For relations involving multiple terms, we normalize by the 2-norm of the largest term, $\Xi=\max_r \|c_r {\bf G}^r\|_2$, where ${\bf G}^r$ is the $r$-th column of $G$. For single-term relations, we instead use the 2-norm of the corresponding tensor before contraction, i.e., $\Xi=\|\nabla {\bf u}\|_2$ for the incompressibility condition \eqref{eq:incomp}.


Parsimonious relations are identified using sequentially thresholded regression (STR). At each step of this iterative algorithm, we compute ${\bf c}$ as the right singular vector of $G$ corresponding to the smallest singular value. This corresponds to the solution of a constrained least squares problem $G^TG{\bf c}=0$ with the normalization $\|{\bf c}\|_2=1$ \cite{gurevich2021}. We start with the full library and, at each step, discard the term $F^r$ with the smallest magnitude, $r=\text{arg\,min} \|c_r {\bf G}^r\|_2$. Note that quality (completeness) of a library can be quantified by computing the residual $\eta$ before any terms are discarded. For data not corrupted by noise, a good library should have $\eta\ll 1$.

After discarding a term from the library, the procedure is repeated, with the residual $\eta_k$  increasing with every iteration $k$ as the number of terms in the model decreases. The iteration is terminated when either there is only one term left or the residual increases by a factor exceeding some threshold, i.e., $\eta_{k+1} > \gamma \eta_k$, where we typically take $\gamma=1.15$. This choice is somewhat arbitrary, as most of our results are robust to a choice $1.1 \leq \gamma \leq 1.3$. In the latter case, we find a multi-term relation, if the final residual is sufficiently small. In the former case, we have to recompute the residual using the normalization appropriate for single-term relations to decide whether that relation is suitably accurate.

Note that STR is not guaranteed to yield the most accurate sparse relation (of a given complexity) contained in the original library. For relatively simple relations (such as the ones identified in this study) and relatively small libraries, we can verify the results of STR by performing a combinatorial search computing the norm of all the relations containing a given number of terms. To validate a relation with $K$ terms contained in a library with $N$ terms through combinatorial search requires an order of $N(N-1)\cdots(N-K+1)$ operations, which is a tractable problem for the values of $K$ and $N$ considered here. For relations with larger $K$, the results of STR can be validated by adding one or more terms from the library that decrease the residual the most. If none of these lower the residual significantly, the result of STR is considered validated.

Multiple relations, including identities, can coexist in the same library. STR is therefore performed iteratively; the library is pruned by throwing out the most complex term of the previously identified relation. STR generally finds identities with machine precision residuals before physical relations with higher residuals. Relations can be more robustly labeled identities by testing them on random smooth synthetic data. Identities will have low residual independent of whether synthetic of experimental data is used, while physical relations are only found for experimental data. Low-dimensional combinatorial searches can more thoroughly explore the relation space to find both identities and physical relations.
We stop searching for new relations once the residual for the full pruned library increases above some threshold e.g. 0.4.

Finally, the coefficients $c_r$ of sparse physical relations are found to vary slightly depending on the number (chosen to be an order of magnitude larger than the size of the library) and location of the integration domains. To quantify the uncertainty in the coefficients, $c_r$ is first identified using the entire feature matrix. The coefficients are then recomputed for the same sparse relation using 100 different samples containing half of the rows in the feature matrix and only the columns corresponding to $c_r$. The mean and standard deviations of the distribution define the value and uncertainty, respectively, of the corresponding coefficient.

\bibliography{biblio}{}
\bibliographystyle{ieeetr} 

\section*{Acknowledgements}
{\bf Funding:} A.F.N and J.N. gratefully acknowledge support from MCIN/AEI/10.13039/501100011033/FEDER,289 (grant No. PID2021-122369NB-100). 

{\bf Competing interests:} All authors declare that they have no competing interests.

{\bf Data and materials availability} All data needed to evaluate the conclusions in the paper are present in the paper and/or the Supplementary Materials. MATLAB code used for this analysis can be found at \begin{verbatim}github.com/mgolden30/SPIDER_active_nematics.\end{verbatim}

{\bf Author contributions:}
Experiment design: A.F.N. 
Experiments and data acquisition: J.N.
Code development: M.G.
Data analysis: M.G., J.N.
Data-driven algorithm development: R.O.G, M.G.
Study conceptualization: A.F.N., R.O.G.
Manuscript preparation: All authors
Manuscript review: All authors

\newpage
\section*{Supplemental material}

\subsection*{Construction of libraries}

Construction of the libraries in an ad-hoc manner is prone to errors, so we developed a systematic procedure. The first step is to construct tensors, up to a certain rank, from the vectors such as ${\bf u}$, ${\bf n}$, and $\nabla$ and scalars such as 1 and $\partial_t$. The number of different such tensors quickly grows with the rank, so we use known physics to further constrain what terms can appear. In particular, the flow is slow, so inertia is negligible. Hence we allow ${\bf u}$ and $\partial_t$ to appear at most once in any tensor. Elastic and viscous effects are described by terms with two spatial derivatives, so we allow $\nabla$ to appear at most twice in any tensor. Lastly, we do not allow mixed spatiotemporal derivatives to appear. There are no physical constraints on ${\bf n}$, so this field can appear an arbitrary number of times. Let us define the fundamental tensors of rank $k$ as $\mathcal{T}_{(k)}$:
\begin{align}
\mathcal{T}_{(0)} & \in\{1\}, \nonumber\\
\mathcal{T}_{(1)} & \in\{ {\bf u}, \,\partial_t {\bf n},\, \partial_t {\bf u}\}\, \cup\, \{{\bf n}\mathcal{T}_{(0)}\} \nonumber\\
\mathcal{T}_{(2)} & \in\{ {\bf u} \partial_t {\bf n},\, \nabla {\bf n},\, \nabla {\bf u}\}\, \cup \, \{{\bf n}\mathcal{T}_{(1)}\}  \nonumber\\
\mathcal{T}_{(3)} & \in\{ {\bf u} \nabla {\bf n},\, \nabla \nabla {\bf n},\, \nabla \nabla {\bf u}\}\, \cup \, \{{\bf n}\mathcal{T}_{(2)}\} \nonumber\\
\mathcal{T}_{(4)} & \in \{(\nabla {\bf n})(\nabla {\bf n}),\,  (\nabla {\bf u}) (\nabla {\bf n}),\, {\bf u} \nabla \nabla {\bf n}\}\,\cup\, \{{\bf n} \mathcal{T}_{(3)}\} \nonumber\\
\mathcal{T}_{(5)}& \in\{{\bf u} (\nabla {\bf n}) (\nabla {\bf n})\}\, \cup\, \{{\bf n} \mathcal{T}_{(4)}\} \nonumber \\
\mathcal{T}_{(k)} & \in \{{\bf n} \mathcal{T}_{(k-1)}\}, \textrm{ for } k>5
\end{align}
The normalization ${\bf n}^2 = 1$ constrains the derivatives of ${\bf n}$: $n_i \nabla_j n_i = 0 $ and $n_i \partial_t n_i = 0.$ A stronger consequence is that the gradient tensor $\nabla{\bf n}$ is reducible \cite{selinger_2018}: $\nabla_i n_j = -n_i b_j + s(\delta_{ij} - n_i n_j)$, where $s = \nabla\cdot{\bf n}$ is the splay scalar and ${\bf b} = -{\bf n}\cdot\nabla {\bf n}$ is the bend vector. The splay $s$ is nematic-covariant, while $b_i$ is nematic-invariant and $n_i b_i = 0.$ This can be used to significantly reduce the number and complexity of the tensors of a given rank. Let us next define the reduced fundamental tensors of rank $k$, $\mathcal{R}_{(k)}$:
\begin{align}
& \mathcal{R}_{(0)} \in \{ 1,\, s,\, s^2\} \nonumber\\
& \mathcal{R}_{(1)} \in \{ {\bf u},\, \partial_t {\bf n},\, \partial_t {\bf u},\, {\bf b},\, s{\bf u},\, \nabla s,\, s{\bf b},\, s^2 {\bf u}\}\,\cup\, \{{\bf n} \mathcal{R}_{(0)}\} \nonumber\\
& \mathcal{R}_{(2)} \in \{ {\bf u} \partial_t {\bf n},\, \nabla {\bf u},\, {\bf u} {\bf b},\, \nabla {\bf b},\, {\bf b}{\bf b},\, s{\bf u}{\bf b},\, s\nabla {\bf u},\, {\bf u} \nabla s\} \,\cup\, \{{\bf n}\mathcal{R}_{(1)}\} \nonumber\\
& \mathcal{R}_{(3)} \in \{ \nabla \nabla {\bf u},\, {\bf u}{\bf b}{\bf b},\, {\bf b} \nabla {\bf u},\, {\bf u} \nabla {\bf b} \} \, \cup \, \{{\bf n} \mathcal{R}_{(2)}\} \nonumber \\
& \mathcal{R}_{(k)} \in \{{\bf n}\mathcal{R}_{(k-1)}\}\textrm{ for } k>3.
\label{eq:tensors}
\end{align}
It is the contractions of these reduced tensors which make up our libraries. This reduction does not remove all identities from the library, but it eliminates many.

\subsection*{Nematic-invariant scalar library}

There are nine nematic-invariant scalars that can be obtained from even-rank reduced fundamantal tensors with the same nematic symmetry listed in \eqref{eq:tensors}:
 \begin{equation}
 F^r \in \{
    1,\, s^2,\,
    n_i u_i s,\, \nabla_i(n_i s),\,
    \nabla_i u_i,\,  \bar{Q}_{ij} \bar{A}_{ij},\,  u_i b_i,\, {\bf b}^2,\, \nabla_i b_i
    \}
    \end{equation}
This library contains a single identity
\begin{align}
    & \nabla_i(n_i s + b_i) = 0
    \label{eq:supp_identity1}
\end{align}
which can be used to prune its last term. These library terms are well-behaved everywhere, and so they can be integrated without any complications yielding the following entries in the feature matrix
\begin{align}
    G^{rkl} = \int_{V_l} w_k F^r\, dV.
\end{align}
Two parsimonious physical relations are identified via symbolic regression, the incompressibility condition \eqref{eq:incomp} and a relation \eqref{eq:stress_balance_scalar} between the director and flow fields
with the relative residuals of $\eta = 0.03$ and $\eta = 0.08$, respectively. Figures S1(a) and S1(b) illustrate how the residual varies with the number $K$ of terms retained in the relation. Note that the coefficient $c_1$ has units of inverse time. Its magnitude is $c_1=O(1)$, which is consistent with our choice of units.

Figure S1(a) shows that there is a version of the incompressibility condition involving 5 terms that has an even lower residual ($\eta=0.02$) than the one-term relation. However, given that our library is missing terms which incorporate the crucial dependence on the microtubule (MT) density $\phi$ (as discussed in the main text), it is rather pointless to look for a physical interpretation of this more general relation.  

\subsection*{Nematic-covariant scalar library}

\begin{figure}
    \begin{tabular}{ccc}
    \centering
    \subfloat[]{
      \includegraphics[width=0.3\textwidth]{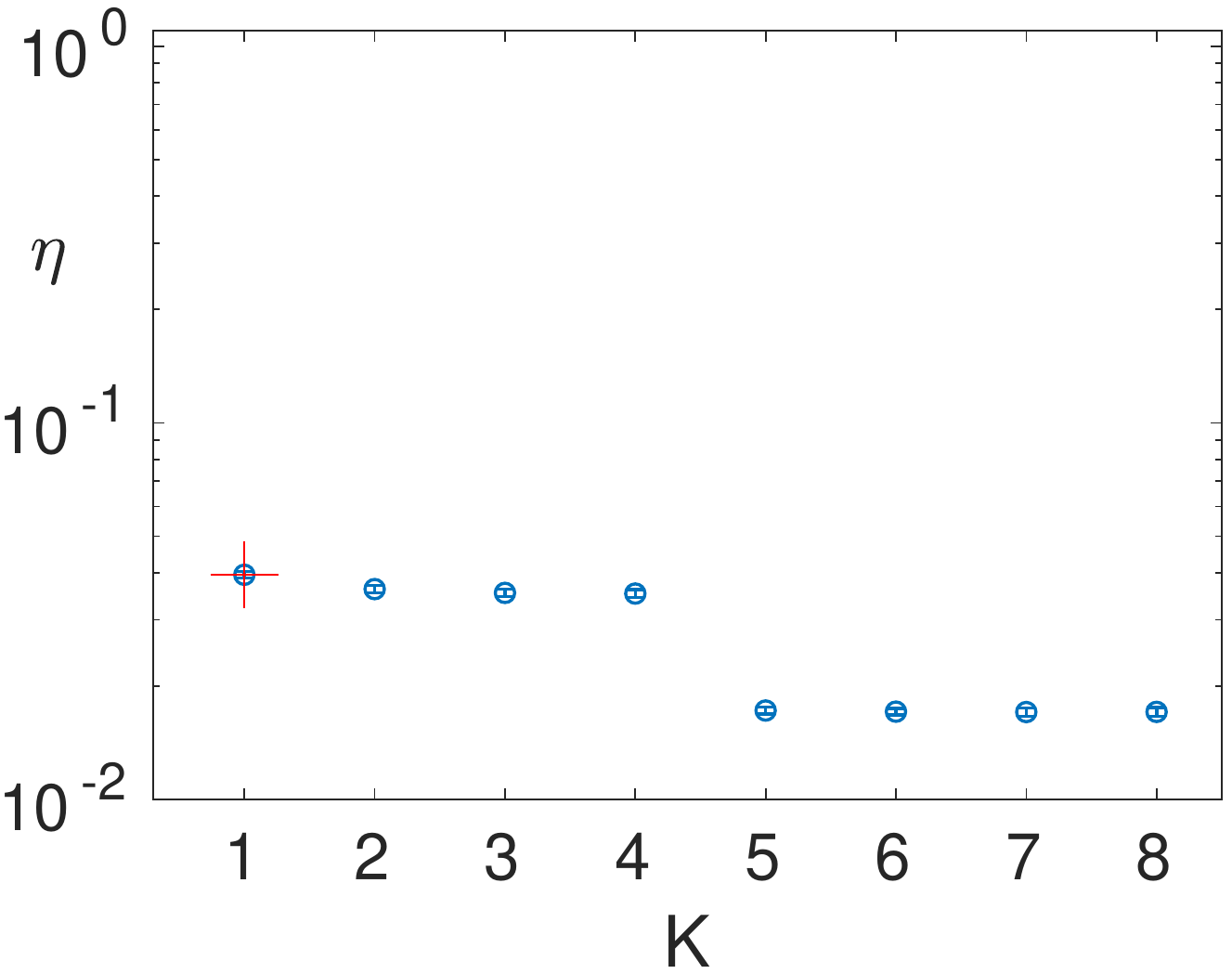}
      \label{fig:histogram}
    }&
    \subfloat[]{
      \includegraphics[width=0.3\textwidth]{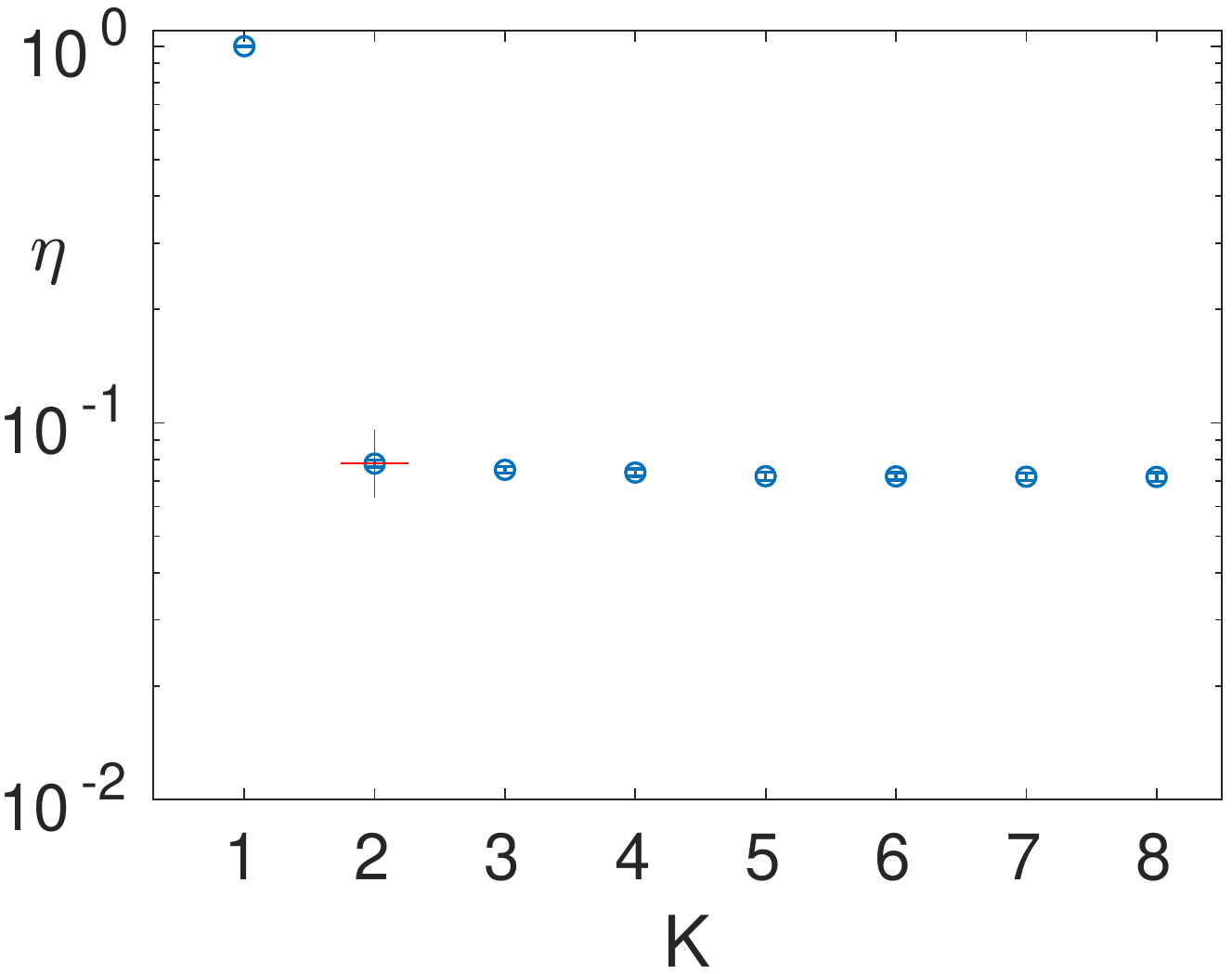}
      \label{fig:histogram}
    }&
    \subfloat[]{
      \includegraphics[width=0.3\textwidth]{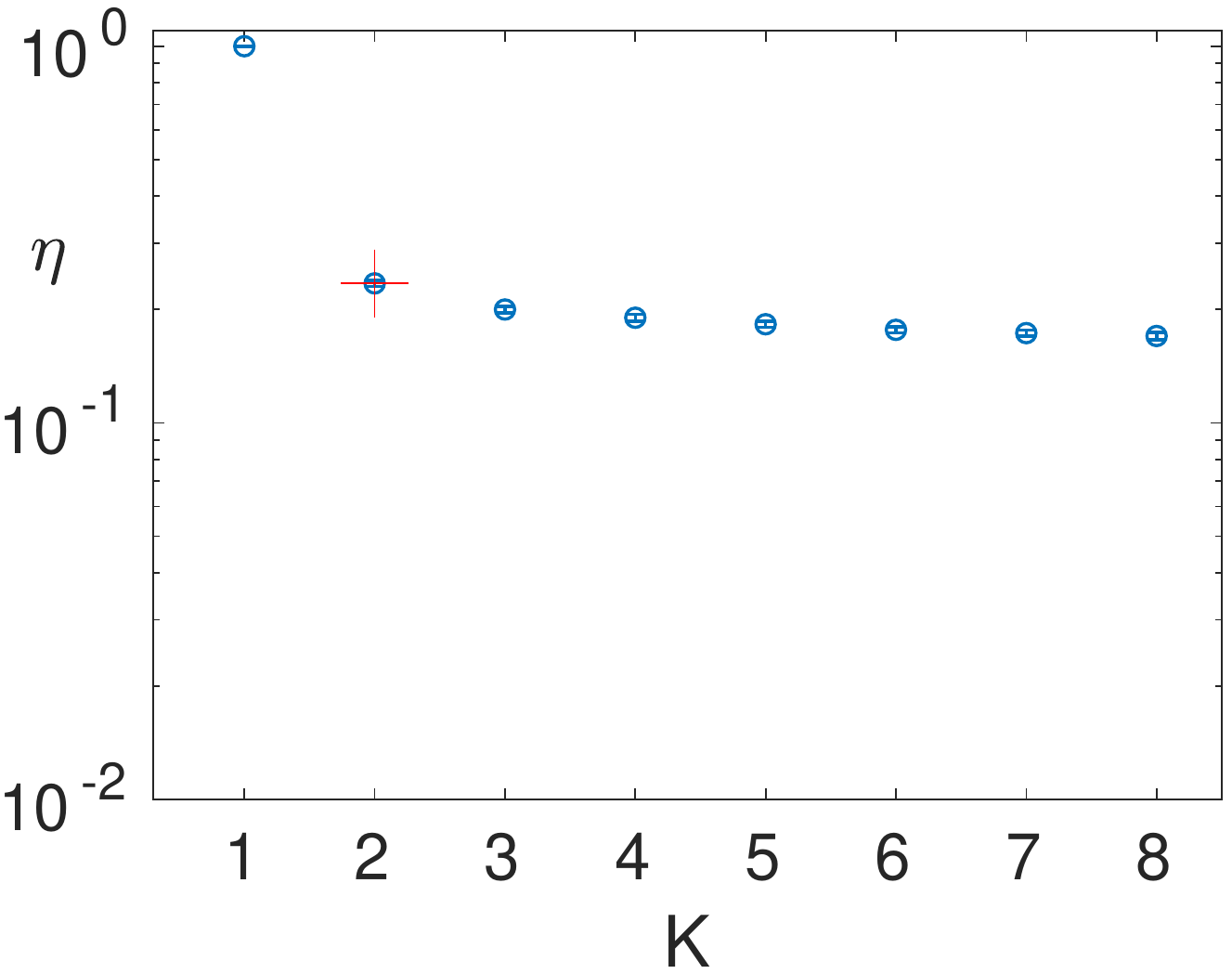}
      \label{fig:histogram}
    }
    \\
    \subfloat[]{
      \includegraphics[width=0.3\textwidth]{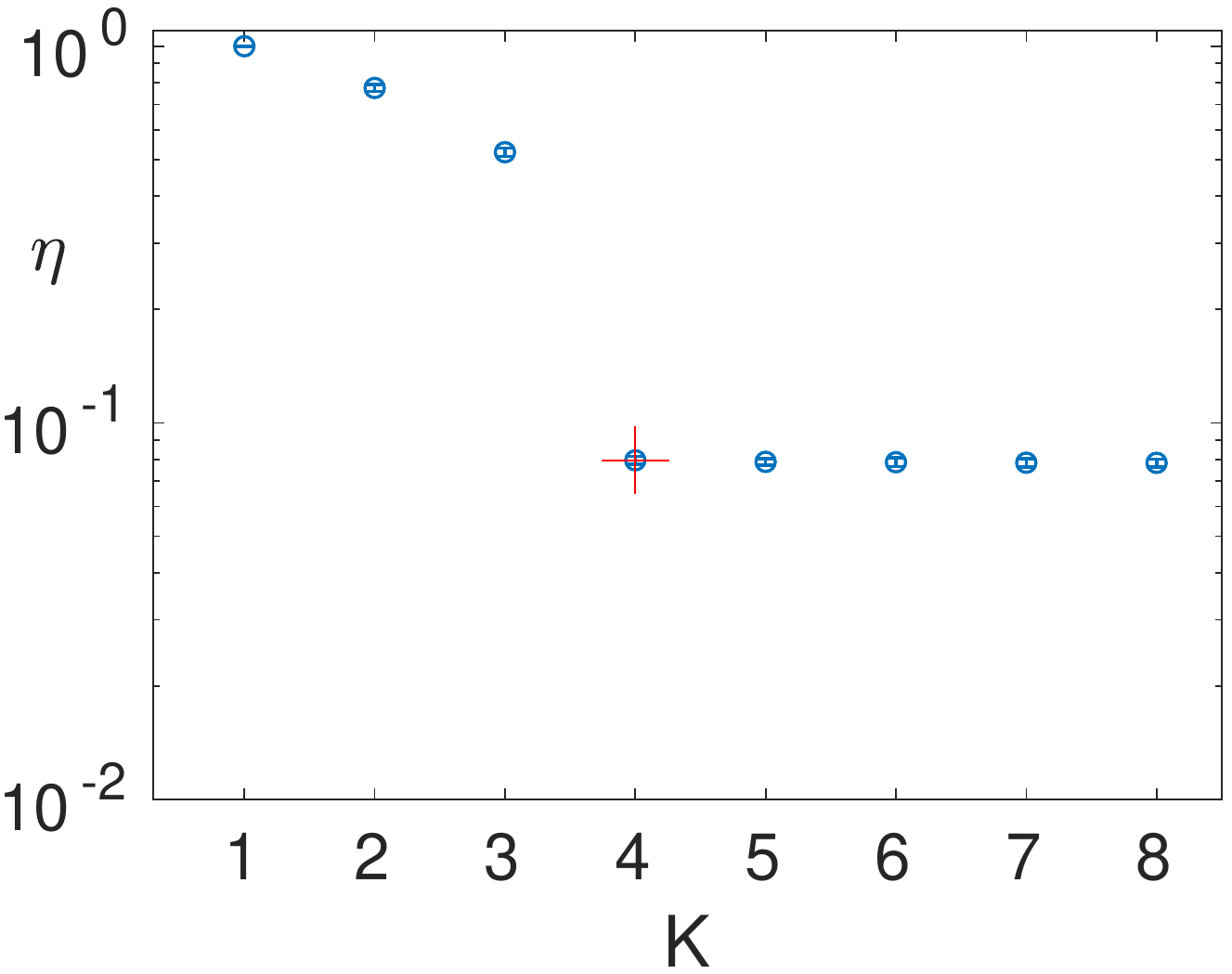}
      \label{fig:histogram}
    }& 
    \subfloat[]{
      \includegraphics[width=0.3\textwidth]{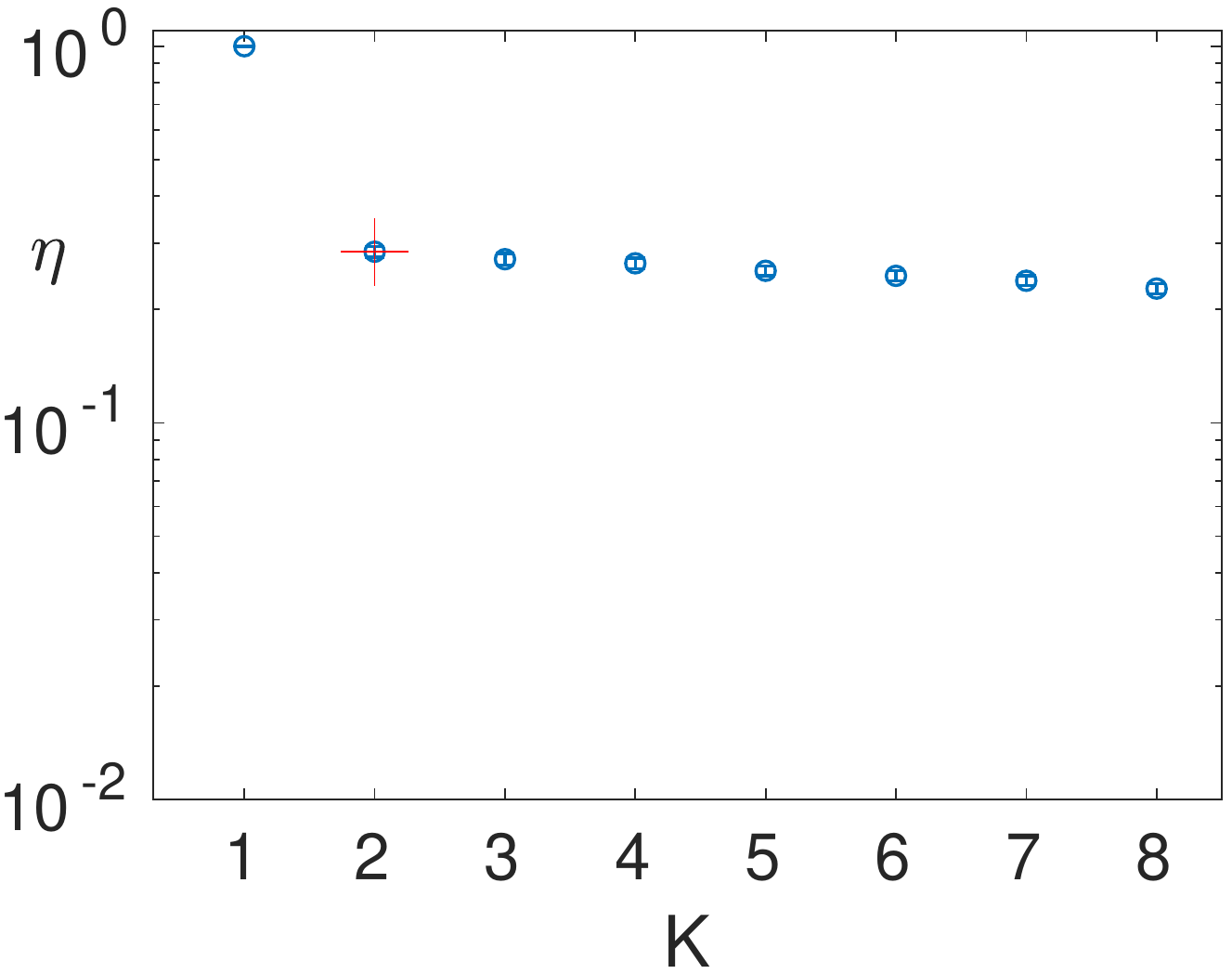}
      \label{fig:histogram}
    }&
    \subfloat[]{
      \includegraphics[width=0.3\textwidth]{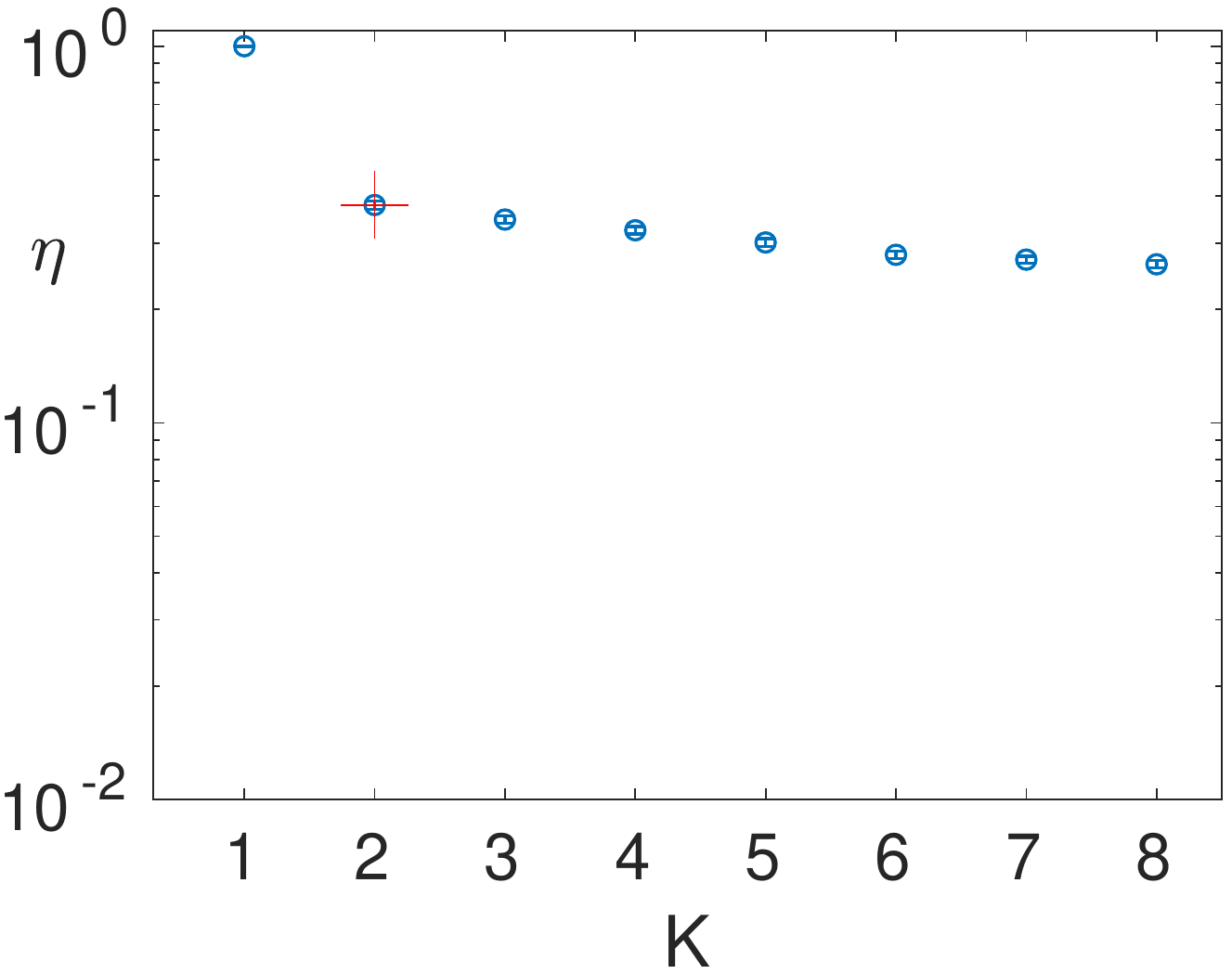}
      \label{fig:histogram}
    }
    \\
    \subfloat[]{
      \includegraphics[width=0.3\textwidth]{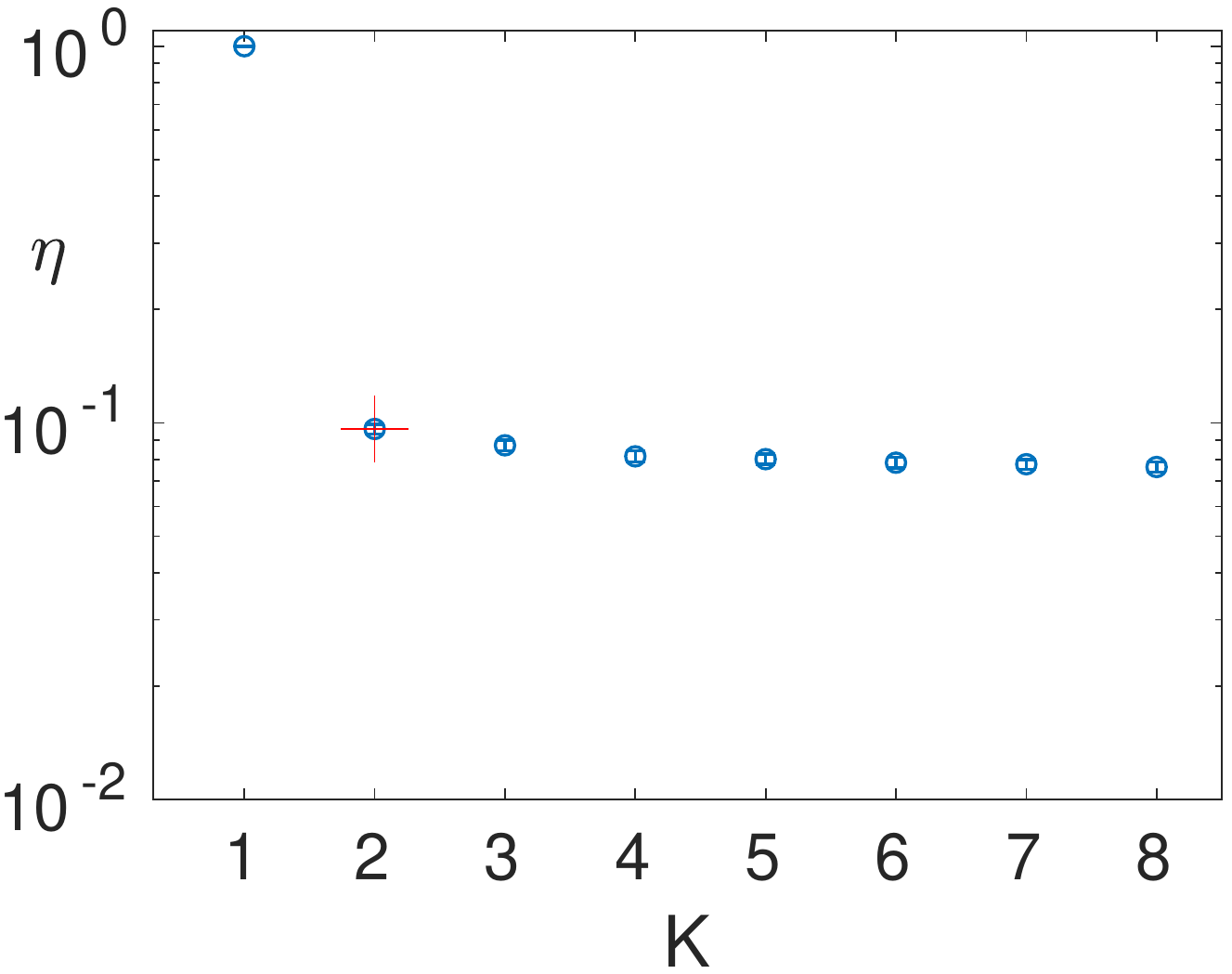}
      \label{fig:histogram}
    }&
    \subfloat[]{
      \includegraphics[width=0.3\textwidth]{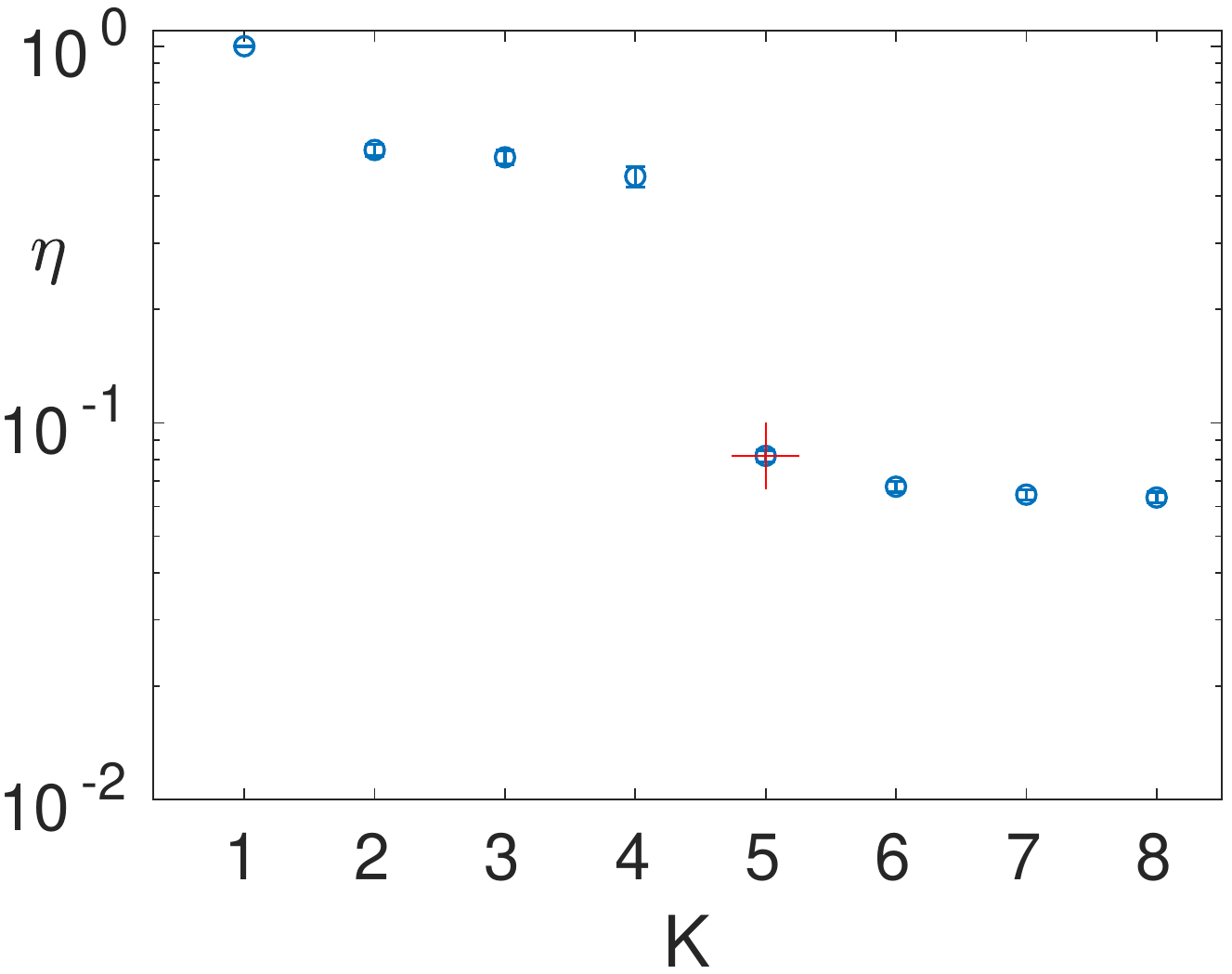}
      \label{fig:supp_STF2}
    }&
    \subfloat[]{
      \includegraphics[width=0.3\textwidth]{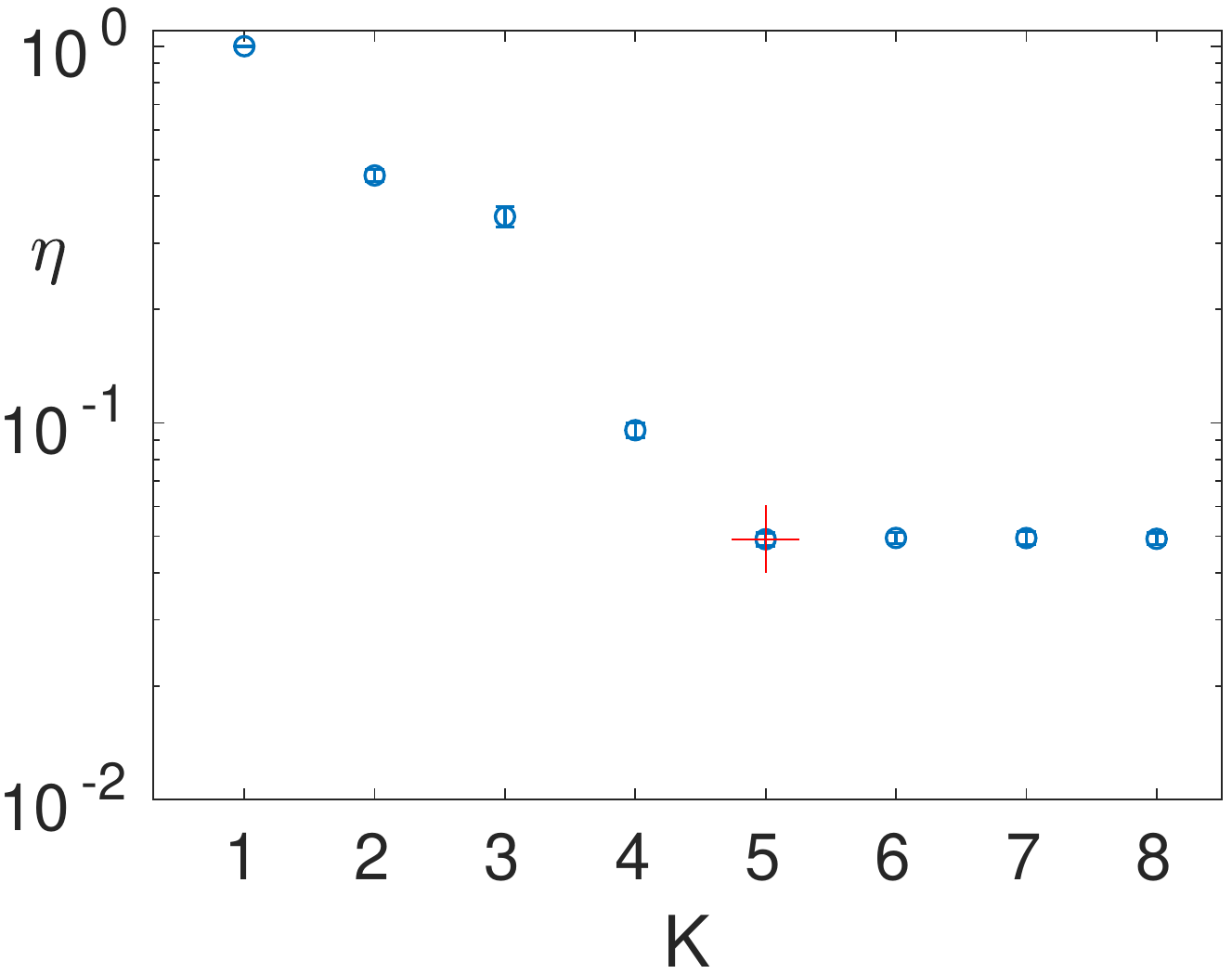}
      \label{fig:supp_antisymm}
    }
    \end{tabular}
    \renewcommand{\figurename}{Figure S1}
    \renewcommand{\thefigure}{}
    \caption{ \textbf{The relative residual $\eta$ as a function of $K$, the number of terms in the relation.} The plots corresponding to discovered equations are 
    (a) equation \eqref{eq:incomp},
    (b) equation \eqref{eq:stress_balance_scalar},
    (c) equation \eqref{eq:stress_balance_scalar_s}, 
    (d) equation \eqref{eq:ndot},
    (e) equation \eqref{eq:supp_stress_balance_vector1}, 
    (f) equation \eqref{eq:supp_stress_balance_vector2}, 
    (g) equation \eqref{eq:stress_balance}, 
    (h) equation \eqref{eq:supp_Qdot}, and
    (i) equation \eqref{eq:supp_antisymm_ndot}. 
    The red cross indicates the corresponding parsimonious relation. These plots were generated by first identifying a parsimonious relations with iterative STR and then adding (removing) terms that decrease (increase) the residual the most (least).}
    \label{fig:sparsification}
\end{figure}

There are 18 nematic-covariant scalars that can be constructed from even-rank reduced fundamental tensors with the same nematic symmetry:
\begin{align}
    \tilde{F}^r \in\ &\{
    s,\,
    n_i u_i,\,
    n_i \partial_t u_i,\,
    u_i \partial_t n_i,\,
    n_i u_i s^2,\,
    s u_i b_i,\,
    s \nabla_i u_i,\,
    s \bar{Q}_{ij} \bar{A}_{ij},\,
    u_i n_i n_j \nabla_j s,\,
    u_i \nabla_i s,\,
    \nonumber
    \\
    &
    n_i n_j n_k \nabla_i \nabla_j u_k,\,
    n_i \nabla^2 u_i,\,
    n_i \nabla_i \nabla_j u_j,\,
    u_i n_i {\bf b}^2,\,
    b_i n_j \bar{A}_{ij},\,
    b_j n_i \Omega_{ij},\,
    u_i n_j \nabla_j b_i,\,
    n_i u_i \nabla_j b_j
    \},
\end{align}
where the last term can be eliminated using the identity \eqref{eq:supp_identity1}.
The nematic-covariant scalar library $\{\hat{F}^r\}$ involves discontinuous fields. Director field changes sign at ``branch cuts'' connecting the topological singularities, causing problems with evaluating derivatives. To restore continuity, all terms are multiplied by ${\bf n}$, making every term a vector. The two components of this vector are treated separately, effectively doubling the number of rows of G (one for $i=1$ and another for $i=2$):
\begin{align}
    \hat{G}^{rkl}_i = \int_{V_l} w_k n_i \hat{F}^r\, dV.
\end{align}
The most accurate parsimonious physical relation identified via symbolic regression contains two terms:
\begin{align}
    & s \left[ \bar{Q}_{ij} \bar{A}_{ij} + c^{(2)}_5 \right] = 0, \label{eq:stress_balance_scalar_s}
\end{align}
where $c^{(2)}_5 = -0.57 \pm 1\%$. This relation follows from \eqref{eq:stress_balance_scalar} as long as $c^{(2)}_5=c'_5$ and, indeed, the values of $c^{(2)}_5$ and $c'_5$ are found to be very close. Figure \eqref{fig:sparsification}(c) shows how the residual varies during regression. The inclusions of the additional factor $s=\nabla\cdot{\bf n}$ increases the residual to $\eta = 0.23$, almost three-fold compared with \eqref{eq:stress_balance_scalar}, which highlights the importance of the quality of the data-processing algorithm (here the one that extracts ${\bf n}$ from the images), especially for relations containing derivatives.

\subsection*{Nematic-covariant vector library}

The nematic-covariant vector library is expected to include an angular momentum balance relation describing the evolution of the director field such as \eqref{eq:LEa} and hence contains the term $\partial_t {\bf n}$. Since $|{\bf n}|=1$, this time derivative should be orthogonal to ${\bf n}$, and without loss of generality, we can restrict our attention to vectors orthogonal to ${\bf n}$ that can be constructed from the reduced fundamental odd-rank tensors with the same nematic symmetry:
\begin{align}
    \tilde{F}^r_i \in \{
    \partial_t n_i,\,
    s u_i,\,
    \nabla_i s,\,
    s b_i,\,
    \bar{A}_{ij} n_j,\,
    \Omega_{ij} n_j,\,
    u_j \nabla_j n_i,\,
    n_j \nabla_j b_i
    \},
\end{align}
where the term $n_j u_j b_i$ has been replaced by its more familiar form $u_j \nabla_j n_i.$ 
We can exclude the component of every term $\hat{\bf F}^r$ along ${\bf n}$ and eliminate the discontinuities in the director field by considering the $z$ component of the vector product $\hat{\bf F}^r\times{\bf n}$ or, in the index notation:
\begin{align}
    G^{rkl} = \int_{V_l} \varepsilon_{ij} w_k n_i \hat{F}^r_j\, dV. \label{eq:feature_vector2}
\end{align}
No identities are found in this library. Symbolic regression identifies one parsimonious physical relation \eqref{eq:ndot}. This relation is formally equivalent to the evolution equation \eqref{eq:LEa} of the Leslie-Ericksen model (sans the elastic contribution $\Gamma{\bf h}$) with the coefficients $c_r$ that are very close to $\pm1$. The relative residual $\eta=0.08$ is quite low and comparable to that of equation \eqref{eq:stress_balance_scalar}. Figure S1(d) shows how it varies during the regression.

\subsection*{Nematic-invariant vector library}

The nematic-invariant vector library would be expected to include a momentum balance relation such as \eqref{eq:LEb}, which contains divergences of various stresses. The library constructed from the reduced fundamental odd-rank tensors with the same nematic symmetry
\begin{align}\label{eq:libv1}
    F^r_i \in\ &\{
    s n_i,\,
    u_i,\,
    (n_j u_j) n_i,\,
    \partial_t u_i,\,
    n_i n_j \partial_t u_j,\,
    b_i,\,
    u_i s^2,\,
    n_i n_j u_j s^2,\,
    n_j u_j \partial_t n_i,\,
    n_i u_j \partial_t n_j,\,
    s n_j u_j b_i,\,
    \nonumber
    \\
    &
    s u_j b_j n_i,\,
    n_i s  \bar{Q}_{jk} \bar{A}_{jk},\,
    s n_j \nabla_i u_j,\,
    s n_j \nabla_j u_i,\,
    s n_i \nabla_j u_j,\,
    n_i (n_j u_j)( n_k \nabla_k s),\,
    u_i n_j \nabla_j s,\,
    \nonumber
    \\
    &
    n_j u_j \nabla_i s,\,
    n_j n_k \nabla_j \nabla_k u_i,\,
    n_j n_k \nabla_i \nabla_j u_k,\,
    u_i {\bf b}^2,\,
    {\bf b}^2 u_j n_j n_i,\,
    u_j b_j b_i,\,
    b_i \bar{A}_{jk} \bar{Q}_{jk},\,
    b_i \nabla_j u_j,\,
    \nonumber
    \\
    &
    b_j \nabla_j u_i,\,
    b_j \nabla_i u_j,\,
    n_j u_j n_k \nabla_k b_i,\,
    u_i \nabla_j b_j,\,
    u_j \nabla_i b_j,\,
    u_j \nabla_j b_i
    \}
\end{align}
will not include stress tensors which involve two spatial derivatives, such as the elastic stress tensor \eqref{eq:elastic_stress_LE}. To get around this, we extended this library by explicitly including higher-order terms of the form $\nabla_j F^r_{ij}$, where
\begin{align}\label{eq:libv2}
F^r_{ij} \in\ &\{ 
    %
    %
    %
    %
    \bar{Q}_{ij},\,
    s^2 \bar{Q}_{ij},\,
    %
    %
    %
    %
    s u_i n_j,\,
    s u_j n_i,\,
    s u_k n_k \bar{Q}_{ij},\,
    (n_k \nabla_k s) \bar{Q}_{ij},\,
    n_i \nabla_j s,\,
    n_j \nabla_i s,\,
    n_i s b_j,\,
    n_j s b_i,\,
    \nonumber
    \\
    &
    %
    %
    (\nabla_k u_k) \bar{Q}_{ij},\,
    (\bar{A}_{kl} \bar{Q}_{kl})\bar{Q}_{ij},\,
    (\nabla_i u_k) n_k n_j,\, 
    (\nabla_j u_k) n_k n_i,\,
    \bar{A}_{ij},\,
    \Omega_{ij},\,
    (u_k b_k) \bar{Q}_{ij},\,
    u_k n_k b_i n_j,\,
    \nonumber
    \\
    &
    u_k n_k b_j n_i,\,
    u_i b_j,\,
    u_j b_i,\,
    (\nabla_k b_k) \bar{Q}_{ij},\,
    (n_k \nabla_k b_i) n_j,\,
    (n_k \nabla_k b_j) n_i,\,
    \nabla_i b_j,\,
    \nabla_j b_i,\,
    {\bf b}^2 \bar{Q}_{ij},\,
    b_i b_j
    \}
\end{align}
is a library of nematic-invariant rank-2 tensors. Note that these tensors include the elastic, viscous and active stresses present in the Leslie-Eriksen model, but not the pressure. Since pressure is a latent field that we have no data for, we should ignore both terms of the form $\nabla_i p$ in \eqref{eq:libv1} and diagonal stress tensors of the form $p \delta_{ij}$ in \eqref{eq:libv2}, where $p$ is any scalar field.. This can be achieved by considering the $z$ component of the vector products $F^r\times\nabla w_k$ in the weak formulation. In the index notation, this corresponds to
\begin{align}
    G^{rkl} = \int_{V_l} \varepsilon_{ij} (\nabla_i w_k) F^r_j\, dV. \label{eq:feature_vector1}
\end{align}

There are a number of identities in this library that will not be listed explicitly. Symbolic regression identified two parsimonious relations:
\begin{align}
    &(\bar{A}_{kl} \bar{Q}_{kl} + c^{(2)}_5 ) s n_i +\nabla_i p = 0, \label{eq:supp_stress_balance_vector1}\\
    &\nabla_k \left[(\bar{A}_{lm}\bar{Q}_{lm} + c^{(3)}_5) \bar{Q}_{ik}  +  \delta_{ik} p\right] = 0, \label{eq:supp_stress_balance_vector2}
\end{align}
where $c^{(2)}_5 = -0.57 \pm 1\%$ and $c^{(3)}_5 = -0.59 \pm 1\%$. Both relations follow from \eqref{eq:stress_balance_scalar} for $c^{(2)}_5 = c'_5$ and $c^{(3)} = c'_5$ and, indeed, these coefficients are found to be rather close. The relative residual of these relations $\eta = 0.28$ and $\eta=0.38$, respectively, are notably higher then the residual for relation \eqref{eq:stress_balance_scalar}, which is not surprising due to the presence of several additional derivatives in the weak form of these relations. The variation of the residuals during regression is shown in Figures S1(e) and S1(f).

\subsection*{Symmetric trace-free tensor library}

It is not always convenient to use a bar to denote the symmetric trace-free part of a tensor. Let us introduce an alternative notation $T_{(ij)} = \frac12\left( T_{ij} + T_{ji} - \delta_{ij} T_{kk} \right)$. We can construct the library of nematic-invariant symmetric trace-free tensors using even-rank reduced fundamental tensors: 
\begin{align}
    \bar{F}^r_{ij} \in\ & \{ 
    %
    %
    %
    %
    \bar{Q}_{ij},\,
    s^2 \bar{Q}_{ij},\,
    %
    %
    %
    %
    \partial_t \bar{Q}_{ij},\,
    s u_{(i} n_{j)},\,
    s u_k n_k \bar{Q}_{ij},\,
    (n_k \nabla_k s) \bar{Q}_{ij},\,
    n_{(i} \nabla_{j)} s,\,
    s n_{(i} b_{j)},\,
    %
    %
    %
    %
    \nonumber
    \\
    &
    \nabla_k u_k \bar{Q}_{ij},\,
    (\bar{A}_{kl} \bar{Q}_{kl})\bar{Q}_{ij},\,
    \bar{A}_{ij},\,
    (u_k b_k) \bar{Q}_{ij},\,
    u_k n_k b_{(i} n_{j)},\,
    (\nabla_k b_k) \bar{Q}_{ij},\,
        \nonumber
    \\
    &
    (n_k \nabla_k b_{(i}) n_{j)},\,
    (\nabla_i b_j)',\,
    {\bf b}^2 \bar{Q}_{(ij)},\,
    \bar{A}_{k(i} \bar{Q}_{j)k},\,
    b_{(i} b_{j)},\,
    u_{(i} b_{j)}
    \}.
\end{align}
This library can be handled in the same way as the nematic-invariant scalar library:
\begin{align}
    &\bar{G}^{rkl}_{ij} = \int_{V_l} w_k \bar{F}^r_{ij}\, dV,\\ 
\end{align}
Note that symmetric trace-free tensors have two independent components $ij=11$ and 12, doubling the number of rows in $G$.

Three identities appear in this library.
\begin{align}
    & \bar{A}_{k(i}\bar{Q}_{j)k} = 0,\nonumber\\
    & b_{(i} b_{j)} + {\bf b}^2 Q_{(ij)} = 0,\nonumber\\
    & u_k \nabla_k Q_{(ij)} - 2s u_{(i} n_{j)}  +  2s u_k n_k Q_{(ij)}  +  2u_k b_k \bar{Q}_{(ij)}  +  2 u_{(i} b_{j)} =0.
\end{align}
We use these to discard the last two library terms. Two physical relations are found in this library, the stress balance relation \eqref{eq:stress_balance} and an evolution equation for the orientation tensor
\begin{align}
    \partial_t Q_{ij} + c'_1 u_k \nabla_k \bar{Q}_{ij}   
    +c'_2 (\Omega_{ik}\bar{Q}_{kj} -\bar{Q}_{ik} \Omega_{kj} )
    +c'_3 \bar{A}_{ij} + c'_4 Q_{ij} (\bar{A}_{kl} \bar{Q}_{kl})=0,
    \label{eq:supp_Qdot}
\end{align}
where $c'_1 = 1 \pm 0.1\%$, $c'_2 = -0.96 \pm 0.1\%$, $c'_3 = -1.02 \pm 0.1\%$, and $c'_4 = 2.05\pm 0.1\%$. These relations have low residuals $\eta=0.1$ and $\eta=0.09$, respectively. For comparison, the tensor balance between $\bar{A}_{ij}$ and $\bar{Q}_{ij}$ proposed in Ref. \cite{thampi2015} has a much higher residual $\eta = 0.67$. The variation of the residuals during regression is shown Figure S1(g-h).
Note that relation \eqref{eq:supp_Qdot} corresponds to the equation \eqref{eq:BEa} of the Beris-Edwards model and can be derived by multiplying the evolution equation \eqref{eq:ndot} by $n_j$. This gives the following correspondence between the coefficients: $c'_1=c_1$, $c'_2=c_2$, $c'_3=c_3$, and $c'_4=-2c_3$.

Using a lower STR threshold $\gamma=1.1$, the residual of the relation \eqref{eq:stress_balance} can be decreased slightly (about 10\%) 
by including a term $A_{ij}$ representing isotropic viscous contribution. The corresponding coefficient is quite small ($-0.02$), which suggests that the isotropic contribution to viscosity is negligible in the regions of high density of MTs. On the other hand, this term is expected to be dominant in the regions with the low density of MTs. Hence, in general, one should expect the viscous stresses to include both contributions ($\bar{Q}_{kl}A_{kl}\bar{Q}_{ij}$ and $A_{ij}$) with the coefficients dependent on the MT density $\phi$. The coefficients are also expected to depend on the shape of the nematic units, similar to the tumbling parameter $\lambda$. It should be possible to derive these coefficients from first principles by solving for the flow above and below the interface with appropriate boundary conditions.

\subsection*{Antisymmetric tensor library}

Let $T_{[ij]} = \frac12\left( T_{ij} - T_{ji} \right)$ be the antisymmetric part of a rank-2 tensor. Again, we can construct the library of antisymmetric tensors using even-rank reduced fundamental tensors: 
\begin{align}
    \tilde{F}^r_{ij} \in\ & \{ 
    %
    %
    %
    %
    %
    %
    %
    %
    n_{[i} \partial_t n_{j]},\,
    s u_{[i} n_{j]},\,
    n_{[i} \nabla_{j]} s,\,
    s n_{[i} b_{j]},\,
    %
    %
    %
    %
    n_k n_{[i} \nabla_{j]} u_k,\, 
    \Omega_{ij},\,
    u_k n_k b_{[i} n_{j]},\,
    u_{[i} b_{j]},\,
    \nonumber
    \\
    &
    n_{[i|} n_k \nabla_k b_{|j]},\,
    \nabla_{[i} b_{j]}
    \}.
\end{align}
This library can be handled in the same way as the nematic-invariant scalar library:
\begin{align} 
    &\tilde{G}^{rkl}_{ij} = \int_{V_l} w_k \tilde{F}^r_{ij}\, dV.
\end{align}
We find a single identity
\begin{align}
    u_k n_k b_{[i} n_{j]}  +  u_{[i} b_{j]} = 0
\end{align}
and a single physical relation
\begin{align}
 n_{[j} \partial_t n_{i]}  +  c''_1 s u_{[i} n_{j]} + c''_2 u_{[i} b_{j]}  + c''_3 \Omega_{ij} + c''_4 n_k n_{[i} \nabla_{j]} u_k  = 0, \label{eq:supp_antisymm_ndot}
\end{align}
where $c''_1 = 1.03 \pm 0.2\%$, $c''_2 = 1.04 \pm 0.9\%$, $c''_3 = -0.98 \pm 0.5\%$, and $c''_4 = -1.00 \pm 0.7\%$. This relation can also be derived from the evolution equation \eqref{eq:ndot} provided $c''_1=-c''_2=c_1$, $2c''_3+c''_4=c_2$, and $c''_4=-c_3$. The relative residual for relation \eqref{eq:supp_antisymm_ndot} is $\eta = 0.05$, making it the most accurate representation of $\partial_t {\bf n}$ found. The variation of the residual during regression is shown in Figure S1(i).

\subsection*{The physical nature of the stress balance relation}

Equations \eqref{eq:stress_balance} and \eqref{eq:stress_balance_scalar} describing the fluid flow can be understood from fist principles in the regions where the curvature of the director field ${\bf n}$ is low. Indeed, MT bundles experience extension in the direction of ${\bf n}$ due to the action of kinesin motors and contraction in the transverse direction due to depletion interaction. The (ATP-concentration-dependent) extension rate $\mathcal{E}$ is the same as the contraction rate to preserve the mean density of MTs at the interface. Let us orient the $x$ and $y$ axes, locally, such that ${\bf n}=\hat{\bf x}$, so that $\bar{Q}_{xx}=-\bar{Q}_{yy}=1/2$ and $\bar{Q}_{xy}=\bar{Q}_{yx}=0$. The kinematic condition at the interface containing the MTs requires a divergence-free interfacial flow to have a velocity
\begin{align}\label{eq:ui}
    {\bf u}^i(x,y)&=\hat{\bf x}\partial_y\psi-\hat{\bf y}\partial_x\psi,
\end{align}
where
\begin{align}
    \psi(x,y)&=\mathcal{E}xy+f(x)+g(y),
\end{align}
is a two-dimensional stream function and $f(x)$ and $g(y)$ are arbitrary functions \textcolorblue{that represent the mean flow. It is easy to see that the flow \eqref{eq:ui} satisfies both equation \eqref{eq:stress_balance} and equation \eqref{eq:stress_balance_scalar} provided $\partial_{xy}\psi=\mathcal{E}=-c_5>0$. Hence the MT bundles are characterized by a constant extension rate $\mathcal{E}$ rather than a constant stress magnitude $\alpha$.}

The corresponding flow above and below the interface can be easily found by assuming its vertical component to vanish everywhere \textcolorblue{(this assumption is clearly invalid in the regions of nonzero $\nabla\cdot{\bf u}$ near topological defects). Let $z=0$ denote the interface between the two fluid layers.} The corresponding flow field in the bottom fluid layer satisfying the kinematic boundary condition at the interface and the no-slip boundary condition at the bottom $z=-h_b$ of the cell is then ${\bf u}^b(x,y,z)=(1+z/h_b){\bf u}^i(x,y)$. Similarly, the flow in the top layer ${\bf u}^t(x,y,z)=(1-z/h_t){\bf u}^i(x,y)$ satisfies the kinematic boundary condition at the interface and the no-slip boundary condition at the top $z=h_t$ of the cell. The corresponding viscous stresses at the interface
\begin{align}
    \sigma_{zx}&=\mu^t\partial_zu^t_x-\mu^b\partial_zu^b_x
    =-\eta{\bf u}^i_x=-\eta[\mathcal{E}x+g'(y)],\nonumber\\
    \sigma_{zy}&=\mu^t\partial_zu^t_y-\mu^b\partial_zu^b_y
    =-\eta{\bf u}^i_y=-\eta[-\mathcal{E}y+f'(x)]
\end{align}
are linear in ${\bf u}_i$ and represent Rayleigh friction \cite{thampi2014,doostmohammadi2016}. 
The friction coefficient
\begin{align}\label{eq:eta}
    \eta=\frac{\mu_t}{h_t}+\frac{\mu_b}{h_b}
\end{align}
depends on the thicknesses of the two layers and their dynamic viscosities $\mu^b$ and $\mu^t$. 

\textcolorblue{We can now obtain a characteristic length scale that balances elastic stresses with viscous stresses caused by the extension of the MT bundles. The initial instability leading to an eventual formation of topological defects and controlling their spacing involves buckling of initially straight filaments \cite{thampi2014epl,martinez2019}. Let us denote the radius and length of our \textcolorblue{MT bundles as $r$ and $L$, respectively.} 
The force exerted by the viscous stresses is given by
\begin{align}
    f_v\sim Lr\frac{\mu\mathcal{E}L}{h},
\end{align}
where $h$ and $\mu$ describe the layer with the higher ratio $\mu/h$, according to the expression \eqref{eq:eta}, and the contribution from the mean flow can be ignored. 
Elastic force at the threshold of the buckling instability is given by
\begin{align}
    f_e\sim\frac{Er^4}{L^2},
\end{align}
where $E$ is the Young's modulus.
Balancing these forces, we find a characteristic length scale
\begin{align}\label{eq:L}
    L\sim r\left[\frac{h}{r}\frac{E}{\mu\mathcal{E}}\right]^{1/4}
\end{align}
which determines the wavelength of the buckling instability.
}

\textcolorblue{
Note that this result yields scaling of the length scale with both the activity, described here in terms of the constant extension rate $\mathcal{E}$, and the viscosity $\mu$ of the fluid which is consistent with experimental observations \cite{lemma2019,martinez2021}. The length scale increases with the stiffness $Er^4$ of the MTs and decreases with the activity ($\mathcal{E}$), as is the case for the standard expression \cite{giomi2014,kumar2018,lemma2019}
\begin{align}
    L\sim\sqrt{\frac{K}{\alpha}},
\end{align}
although the functional dependence is clearly different. 
Given that in our experiments, $\mu\sim 10^{-3}$ Pa$\,$s, $h\sim 50$ $\mu$m, and $\mathcal{E}\sim 0.015$ s$^{-1}$, and using the known values of $r$ and $E$ for individual MTs, $r\sim 25$ nm and $E\sim 10^8$ Pa \cite{kis_2002, zhang_guan_2020}, we obtain $L\sim 270$ $\mu$m which is in good agreement with the mean spacing of $\sim$240 $\mu$m between same-charge defects.
}

\section*{Supplemental Movie Captions}

Movie S1: \textbf{The divergence of the interfacial flow.} The divergence of the experimental flow field (left) is compared with the corresponding experimental images (right). The scale of the color bar is arbitrary, and the black contours correspond to boundaries of low-density regions.
\\

\noindent
Movie S2: \textbf{The angular velocity of the microtubules}. The left panel shows the angular velocity $\partial_t \theta$ obtained by finite differencing the data, and the right panel shows the reconstruction using the right-hand side of the discovered PDE (3). The scale of the color bar is arbitrary, and the black contours correspond to boundaries of  low-density regions.
\\

\noindent
Movie S3: \textbf{Comparison of the two terms in the effective stress balance equation (4).} A diagonal component of the active stress tensor is shown on the left and the same component of the viscous stress tensor is shown on the right. The scale of the color bar is arbitrary, and the black contours correspond to boundaries of  low-density regions.
\\

\noindent
Movie S4: \textbf{The mask used in analyzing the data.} The mask used to filter out data from low-density regions in our analysis (left) is compared with the corresponding experimental image (right). The mask (red) is overlaid with the  director field (black arrows).

\end{document}